# A 0.4 V, 19 pW Subthreshold Voltage Reference Generator Using Separate Line Sensitivity and Temperature Coefficient Correction Stages

**Mohammad Azimi, Mehdi Habibi*, Hamid Reza Karimi-Alavijeh**

**Department of Electrical Engineering, Sensors and Interfaces Research Group, University of Isfahan, Isfahan, Iran. *Corresponding Author, Email: mhabibi@eng.ui.ac.ir**

**Abstract-** Sensor nodes and IoT systems require blocks that not only consume low power but also have good accuracy. Voltage reference generators are also considered important building blocks in sensor interface circuits. This paper presents a solution to increase the accuracy of low power subthreshold voltage generators by lowering the circuit sensitivity to temperature and supply voltage variations. The enhancement is achieved by using two separate stages for temperature coefficient (TC) and line sensitivity (LS) correction. A 0.18 µm standard CMOS process has been used for the proposed structure. The effects of parameter variations in the fabrication process are investigated using post-layout simulation and Monte Carlo analysis. In the supply voltage of 0.4 V to 2 V, an LS of 143.8 ppm/V is obtained. In typical corner conditions, the achieved TC is 7.45 ppm/°C over the temperature range of 0 °C to 80 °C. Due to process changes, and mainly affected by threshold voltage variations, the average TC can change to 39.2 ppm/°C. The minimum power consumption at 0 ºC and at a supply voltage of 0.4 V is 3.25 pW while the power consumption increases to 2.84 nW in 80 ºC and at the maximum supply voltage of 2 V.

## 1-Introduction

Voltage reference generators are one of the principle blocks used in many analog/mixed signal circuits such as data converters [1], sensors [2], and regulators [3]. With the development of applications such as the IoT and energy harvesting circuits in which power consumption is important [4], sub-threshold voltage references consuming picowatt power have attained more attention. Voltage reference generators with output Independent of Temperature (IT) can be designed by combining circuits which produce output voltage or currents Proportional To Absolute Temperature (PTAT) and Complementary To Absolute Temperature (CTAT) [5,6]. Traditionally, bandgap voltage references are one of the most common circuits for implementing voltage references. Despite the possibility of high-order temperature curvature compensation and the good accuracy of these structures, the minimum voltage and power consumption are among the problems of these circuits [7,8]. Replacing BJT transistors in previous designs with MOSFETs and introducing all-MOSFET solutions can reduce the voltage and power consumption of these circuits. However, process-dependent changes in the threshold voltage have reduced the accuracy of these circuits compared to BGRs [9]. Among all-MOSFET voltage reference generators with low voltage and power consumption, structures with one or two active loads that operate in the sub-threshold region are common [10–16], as shown in Figure 1 [17].

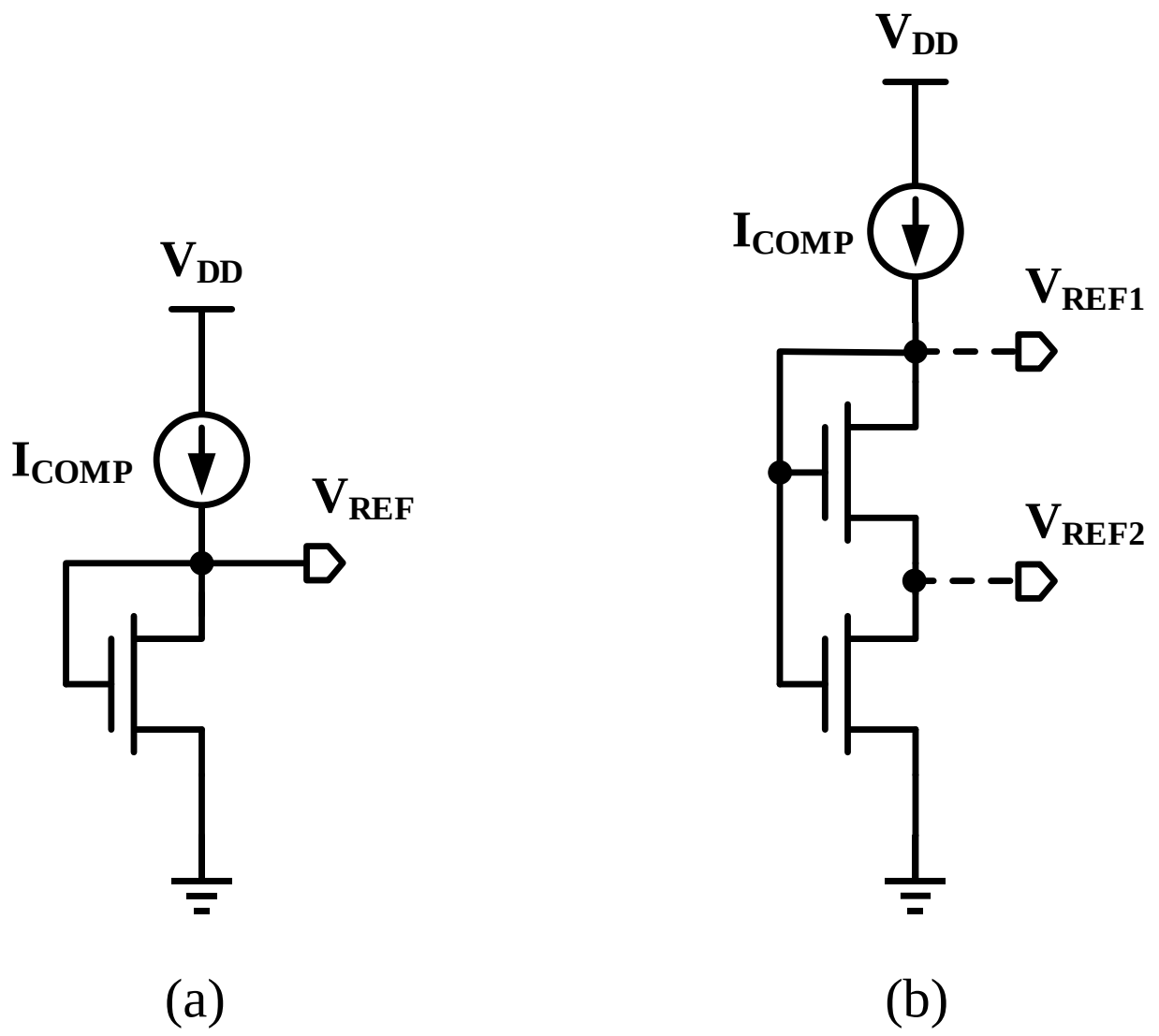

Figure 1. Basic structures to generate the reference voltage using (a) one active load (b) two active loads

In these structures, the temperature compensation current ($I_{COMP}$) can be implemented with a single transistor or more complex circuits [18]. The generated reference output voltage is based on the threshold voltage difference of the transistors. The circuit proposed in [19] uses a complex current generating circuit which increases its power consumption. A two transistor voltage reference generator is presented in [11] which uses a native transistor as a current source with its gate connected to ground. The minimum power supply is 0.5 V and the obtained TC is 62 ppm/°C without using a trim circuit. The use of a native transistor as a current source results in a suitable LS of 0.033%/V.

In [12], by replacing the current source transistor with a low threshold voltage device and connecting its gate to source, and also using a high threshold voltage transistor as the load, a new structure is proposed that can operate with a minimum supply voltage of 0.45 V while its TC and LS are increased to 105.4 ppm/°C and 0.46%/V respectively. To further reduce the supply voltage, it is necessary to use similar type transistors for the current source and the load. Furthermore the gate voltage of the upper transistor should be reduced. With these modifications, the minimum operable power supply is reduced to 0.15 V in [13]. Although this circuit has a low supply voltage and power consumption, but it has a TC and LS of 1462.4 ppm/°C and 2.03%/V respectively which is much higher than the circuits given in [11] and [12].

Using a feedback path and presenting a two stage structure, the reference generator given in [14] improves the two TC and LS factors to 246.6 ppm/°C and 0.28%/V respectively, although it requires a higher minimum power supply of 0.2 V. In [15], using the two transistor active load structure and a native transistor similar to [11] as the current source device, a reference voltage generator is constructed. The generated output voltage is dependent on the threshold voltage difference between the load transistors. In addition, it is shown that the PSRR value improves as the number of current source transistors increases. Subsequently, the circuit is modified using a two transistors current source and the PSRR is reduced to -98 dB at 100 Hz. Higher values of PSRR are observed at higher frequencies. The voltage reference generator in [16] is similar to [15] but with a different current source structure. In this architecture, the current source is biased using the reference voltage and then mirrored back to the active loads. Using a trim circuit at the output, a TC value of 72.4ppm/℃ is achieved at a minimum power supply voltage of 0.45V. The obtained LS is 0.15%/V. Various methods have been proposed to improve the LS, including the use of cascode structures [16,20], cascode current mirror [21], and long-channel transistors. Despite the effectiveness of these methods, the use of cascode structure increases the minimum required supply potential, while increasing the channel length is mostly effective at low channel lengths. In [22] a structure is given to compensate the DIBL effect of [16] and improve LS from 0.15%/V to 0.019%/V.

In this paper, a low power voltage reference circuit with a suitable TC level is presented by controlling the voltage of the current source transistor gate in a one active load structure. At the same time, by using the two-stage technique of [14] where the output of a first stage is used to drive the power supply of a second stage and also presenting a DIBL compensating structure, the LS of the proposed circuit is reduced to a relatively low level.

The rest of the paper is organized as follows: the TC correction stage is described in Section 2 and circuit analyses for temperature compensation are presented. In section 3, the proposed voltage reference is completed by adding the LS correction stage, and also the equations related to LS are extracted. In section 4, the PSRR analyses are presented. Section 5 introduces a trimming circuit to compensate for variations in TC between different process corners. Section 6 discusses the optimal aspect ratio for temperature compensation. Section 7 describes the dimensions and layout design of the transistors. Section 8 presents simulation results and compares them to recent all-MOSFET low power reference generators. In the final section, some concluding remarks are given.

# 2- Temperature compensation

The general proposed circuit is shown in Figure 2 to generate a voltage reference output with low TC. $M_3$ is a thick oxide NMOS while $M_4$ is a thin oxide transistor. Thick oxide transistors are shown with a thicker gate line. Since the circuit operates in the subthreshold region, the subthreshold current is given as follows [13,23]:

$$I_D = \mu_n C_{ox} \frac{W}{L} V_T^2 exp\left(\frac{V_{GS} - V_{TH}}{nV_T} + \delta V_{GD}\right)\left(1 - exp\frac{-V_{DS}}{V_T}\right) \tag{1}$$

where $\mu_n$ is the electron mobility, $C_{ox}$ is the oxide capacitor per unit area, $W$ and $L$ are the width and length of a transistor. $V_T = k_B T/q$ is the thermal voltage ($k_B$ is Boltzmann's constant, $T$ is the absolute temperature and $q$ is the electron charge), $V_{TH}$ is the threshold voltage, $n = 1 + C_d/C_{ox}$, and δ is the GIDL factor [23,24]. $V_{GS}$ and $V_{DS}$ are the gate-source and drain-source voltages, respectively. The subthreshold current is not affected by channel length modulation. With $V_{DS}$>4$V_T$ and neglecting the effect of GIDL, which is not significant in the subthreshold region [11], in Figure 2, $I_{D,4}$ can be simplified to

$$I_{D.4} = \mu_n C_{ox.L} (\frac{W}{L})_4 (\frac{k_B.T}{q})^2 exp\left(\frac{V_{GS.4} - V_{TH.4}}{n_L V_T}\right) \tag{2}$$

where $V_{GS,4}= V_{G4}-V_{REF}$, while $n_L$ and $C_{ox,L}$ are subthreshold slope factor and oxide capacitance for thin oxide transistor, respectively. Similarly, $V_{REF}$ can be expressed using the subthreshold equation as follows:

$$V_{REF} = V_{TH.3} + n_H V_T \ln\left(\frac{q^2 I_{D.4}}{\mu_n C_{ox.H}(W/L)_3 k_B^2 T^2}\right) \tag{3}$$

where $n_H$ and $C_{ox,H}$ are subthreshold slope factor and oxide capacitance for thick oxide transistor, respectively. To express the concept, assuming $n = n_H = n_L$, $(W/L)_3=(W/L)_4$, and substituting $I_{D,4}$ from (2) in (3), the $\mu T^2$ term is cancelled out in the "ln" function and the output reference voltage $V_{REF}$ can be expressed with (4) as follows:

$$V_{REF} = V_{TH.3} - V_{TH.4} + V_{G.4} + nV_T \ln\left(\frac{C_{ox.L}}{C_{ox.H}}\right) - V_{REF} \tag{4}$$

$$\rightarrow\ V_{REF} = \frac{V_{TH.3} - V_{TH.4} + V_{G.4} + nV_T \ln\left(\frac{C_{ox.L}}{C_{ox.H}}\right)}{2}$$

$V_{TH3}$ and $V_{TH4}$ are shown in Figure 3(a) as a function of temperature. The TC of the thick oxide transistor ($M_3$) threshold voltage is about -1mV/°C while for the thin oxide it is about -0.75 mV/°C. Also $V_{TH,3}$-$V_{TH,4}$ is plotted against temperature in Figure 3(b). $V_{TH,3}$-$V_{TH,4}$ exhibits CTAT behavior and the TC of $nV_T ln(C_{ox,L}/C_{ox,H})$ term is lower than that of $V_{TH,3}$-$V_{TH,4}$. Therefore, the gate voltage of $M_4$, $V_{G,4}$, in Figure 2 should be controlled such that the CTAT behaviour is cancelled out using $V_{G,4}$ [10]. In the following it will be shown that a PTAT voltage generator on the gate terminal of $M_4$ can compensate these threshold voltage changes.

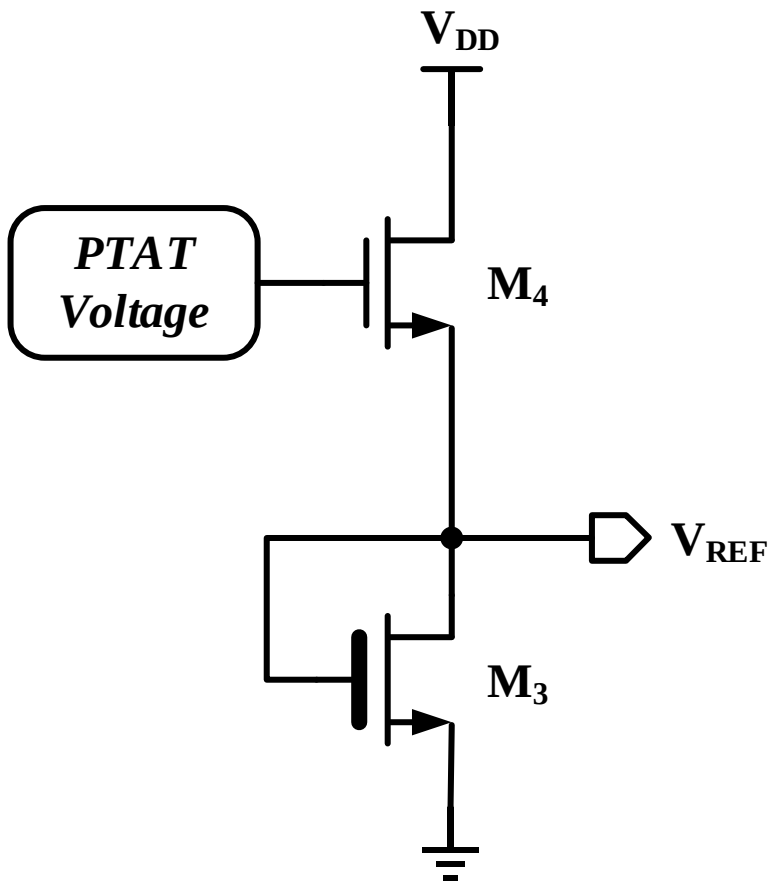

Figure 2. The proposed scheme for temperature compensation

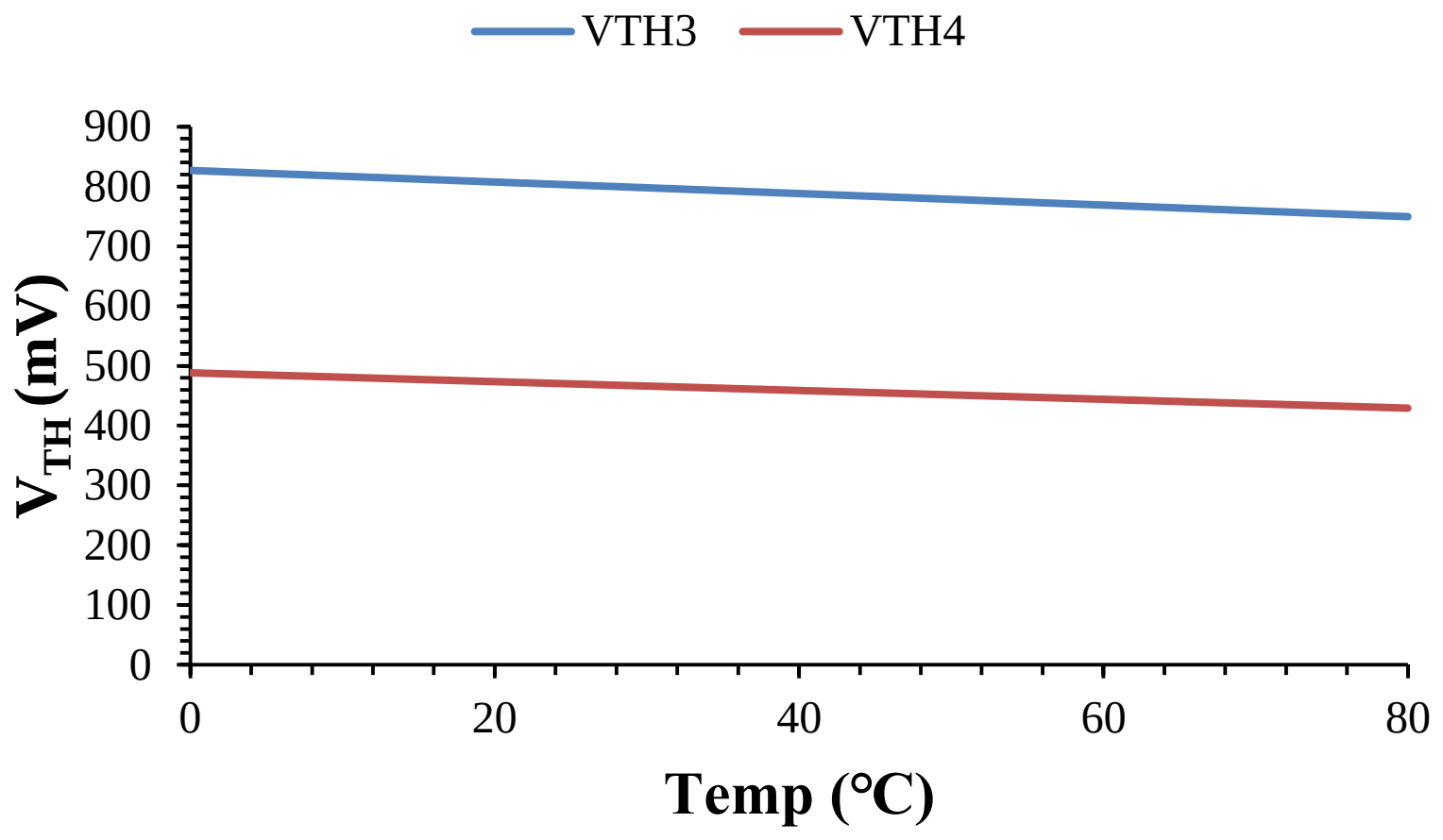

(a)

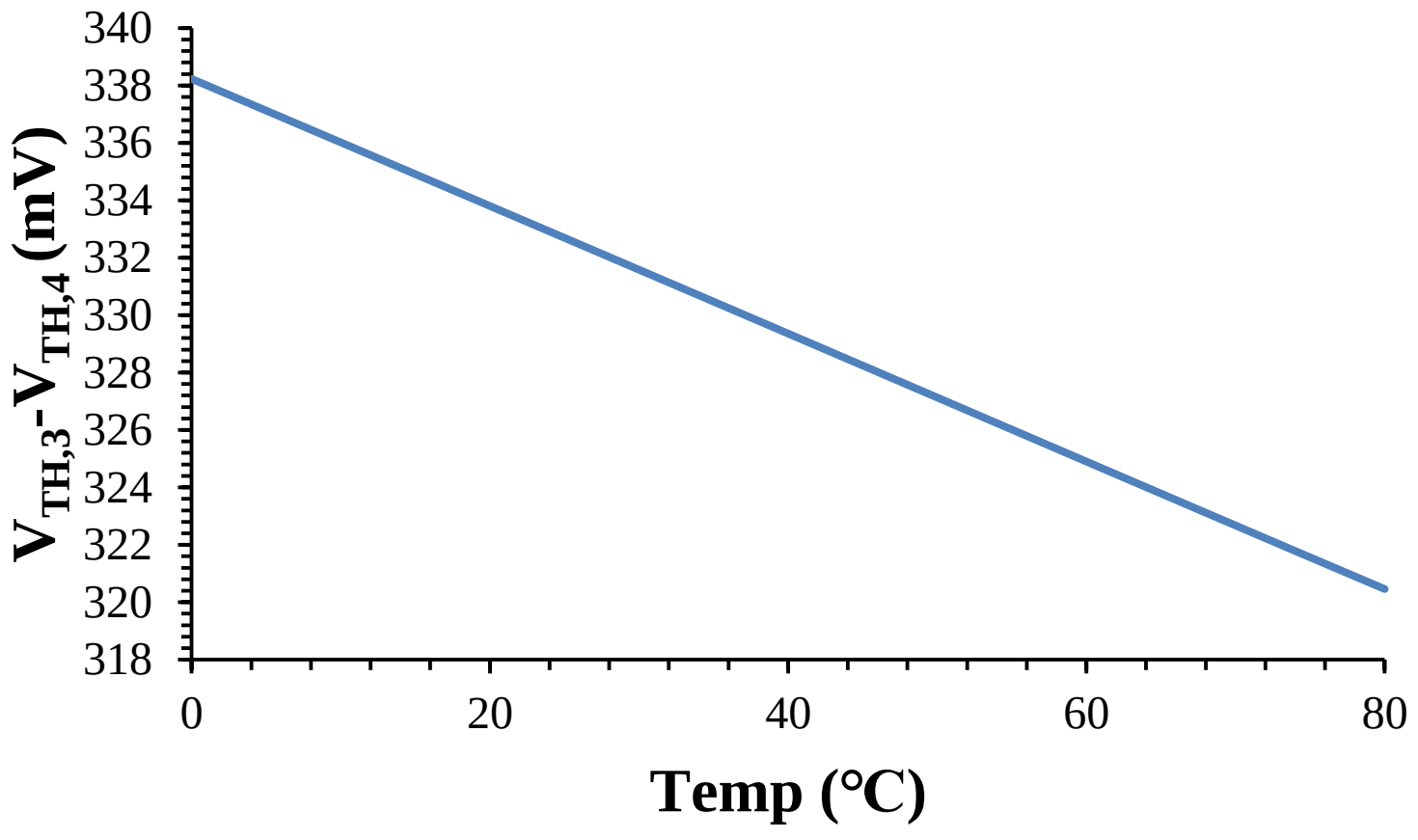

(b)

Figure 3. Variations of (a) $V_{TH3}$ and $V_{TH4}$ (b) $V_{TH3}$-$V_{TH4}$ versus temperature from 0 °C to 80 °C

As shown in Figure 4, the structure presented in [14] is used to implement the PTAT voltage. $M_1$ and $M_2$ are thin oxide transistors. By controlling the current injection from the current source and adjusting the load transistor dimensions, it is possible to produce PTAT voltages with different slopes against temperature in both the single load and two-load structures [25]. However, one active load structure is used in this paper to reduce the supply voltage. The aspect ratio of transistors $M_1$ and $M_2$ may be adjusted in this structure so as to generate PTAT, CTAT, and IT voltage references. Although the transistors are similar, however Narrow Width Effect (NWE) and Short Channel Effect (SCE) can be used to create different threshold voltages. In Figure 5, the output voltage versus temperature variations for the various aspect ratios $R = (W/L)_2/(W/L)_1$ are examined. For three different aspect ratios of R(1) = 0.4, R(2) = 0.8 and R(3) = 1.6, the CTAT, IT and PTAT Voltages are obtained, respectively. Note that the value of the IT output voltage according to Figure 6 is approximately equal to the threshold voltage difference, $\Delta V_{TH}$, between the two transistors. $M_1$ and $M_2$ have threshold voltages of 437 mV and 406 mV, respectively, at 25 °C. Therefore, their difference is 31 mV, close to the output voltage of 38 mV.

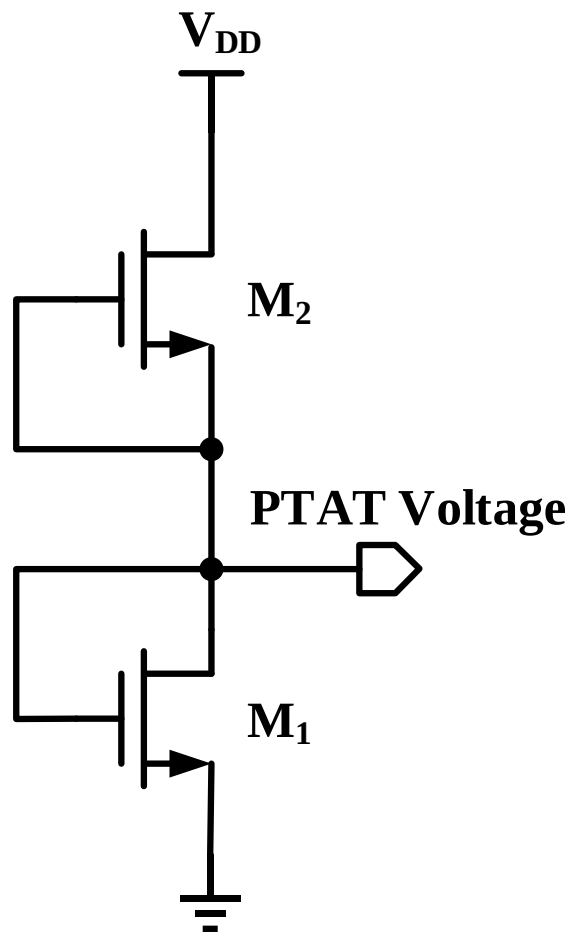

Figure 4. PTAT/CTAT or IT generator [14]

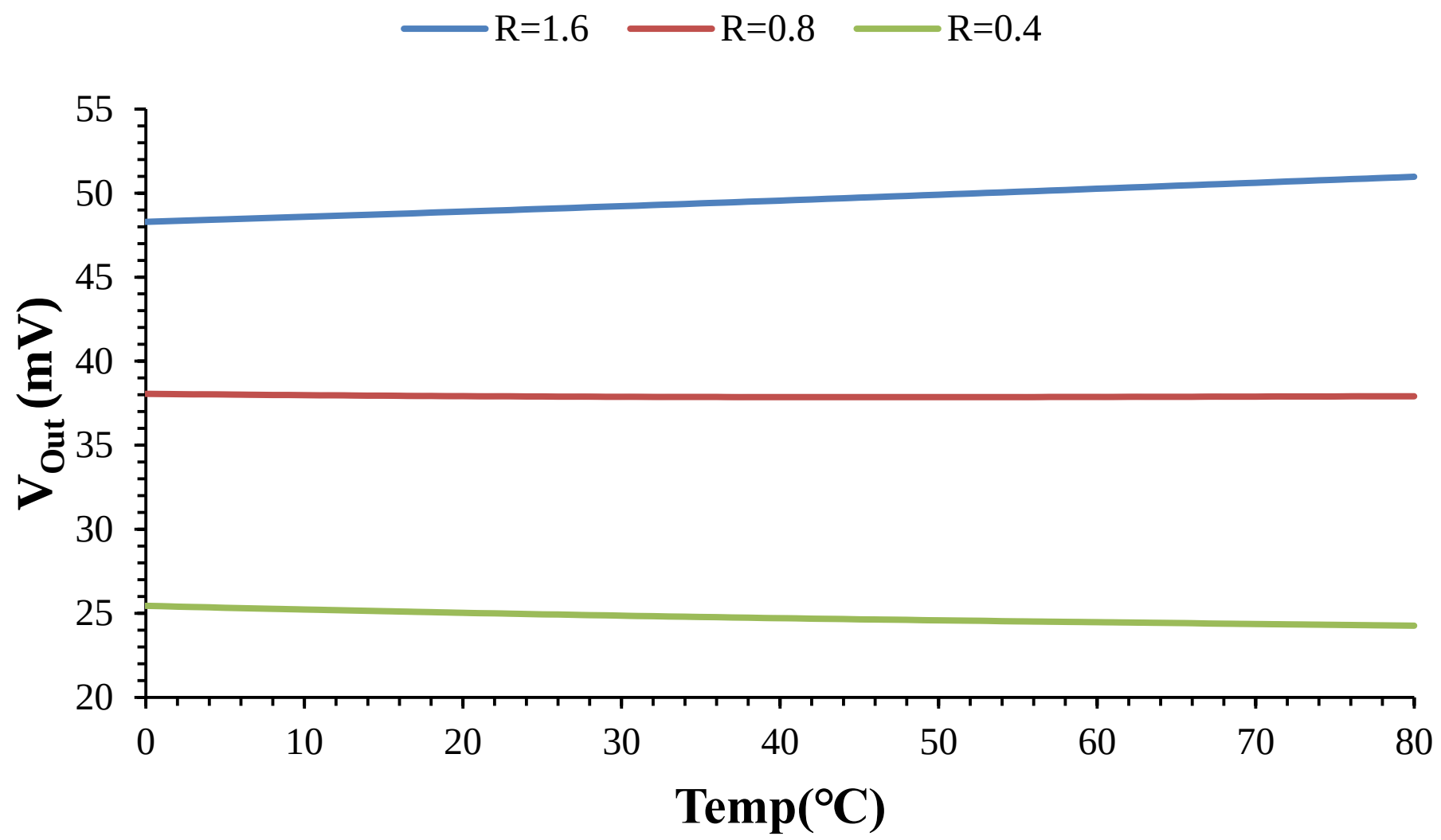

Figure 5. Output voltage versus temperature for different aspect ratios R(1)=0.4, R(2)=0.8, and R(3)=1.6

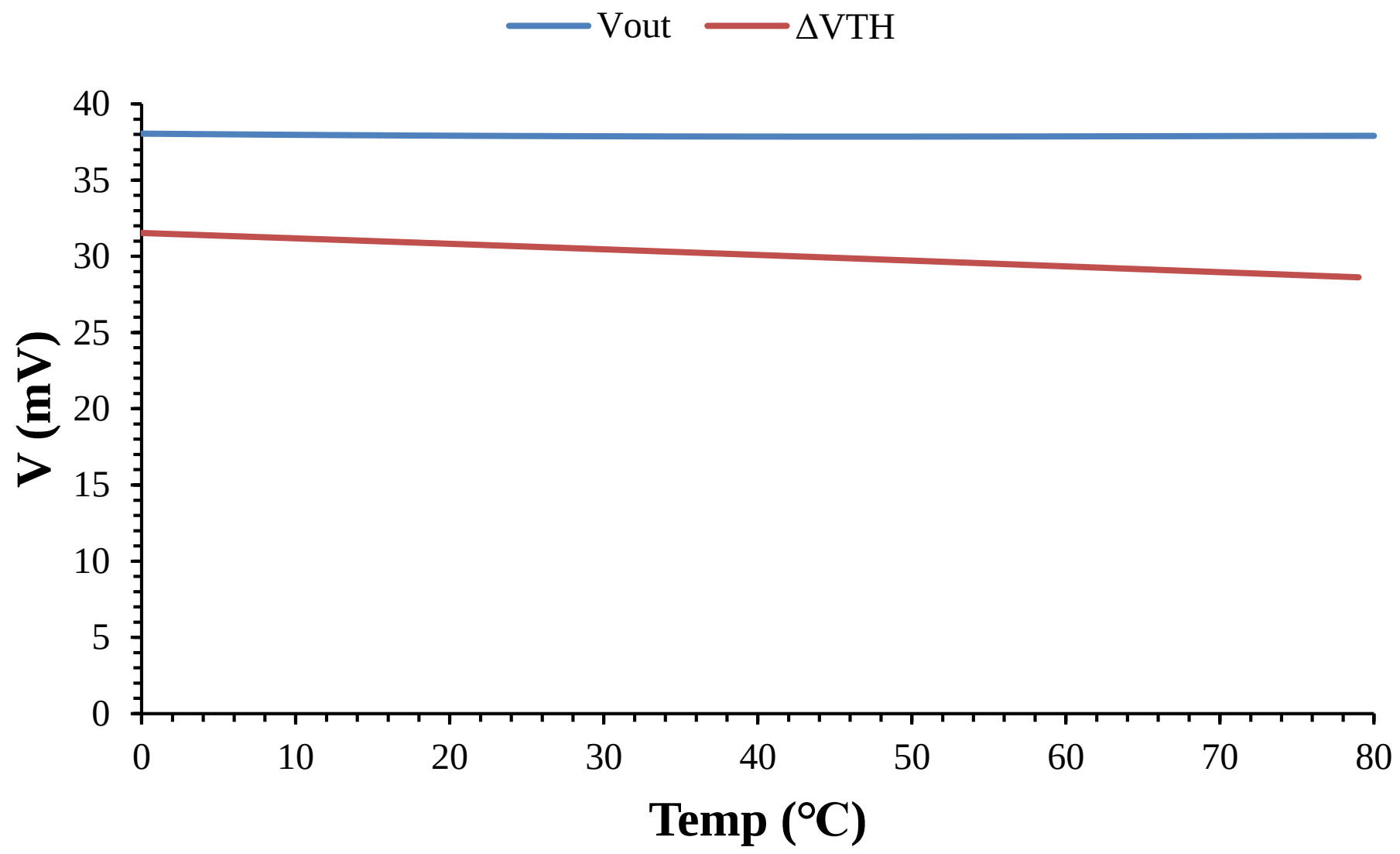

Figure 6. Output voltage and $\Delta V_{TH}$ versus temperature from 0 ºC to 80 ºC

Based on Figure 2 and 4, the proposed temperature compensation circuit is shown in Figure 7. The output voltage equation is obtained in the following by considering appropriate temperature compensation requirements. PTAT voltage is obtained using $M_1$ and $M_2$ drain currents according to Equation (5).

$$V_{PTAT} = V_{TH.1} - V_{TH.2} + n_L V_T \ln(\frac{(W/L)_2}{(W/L)_1}) \quad (5)$$

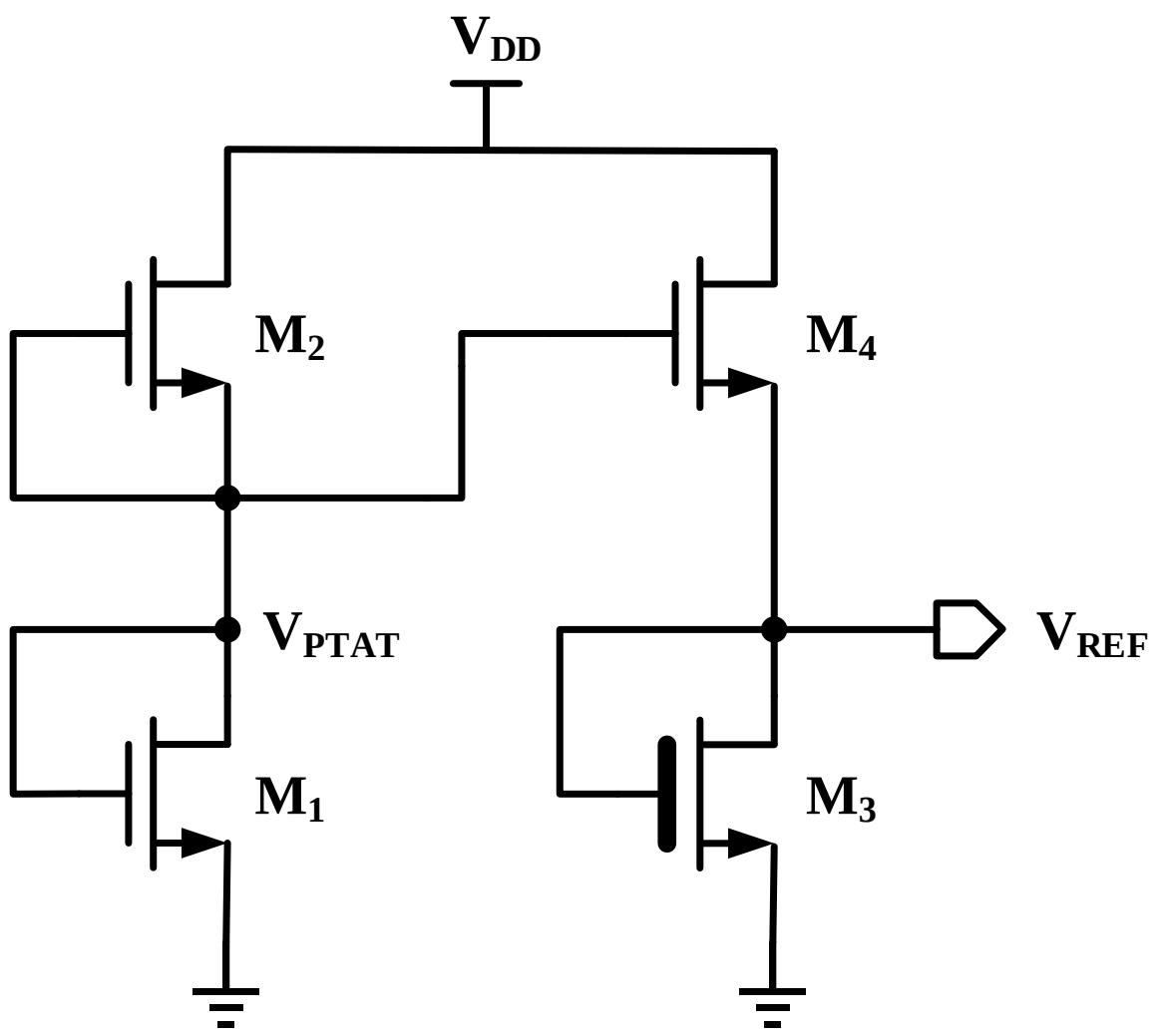

Figure 7. The proposed circuit for temperature compensation

For large enough $(W/L)_2/(W/L)_1$ ratios, the last term of equation (5) will be PTAT and the dominant term.

Using (5), the output voltage can be determined as follows:

$$V_{REF} = \frac{n_L V_{TH.3} + n_H(V_{TH.1} - V_{TH.2} - V_{TH.4}) + n_H n_L V_T \ln\left(\frac{C_{ox.L}}{C_{ox.H}} \frac{(W/L)_4}{(W/L)_3} \frac{(W/L)_2}{(W/L)_1}\right)}{n_H + n_L} \quad (6)$$

The threshold voltage and thermal voltage are functions of temperature, while the subthreshold swing factor weakly depends on temperature [22]. The threshold voltage is a complex function of temperature, and the following expression is used to describe $V_{TH}$ using first-order approximation [26]:

$$V_{TH} = V_{TH}(T_0) + \alpha(T - T_0) \quad (7)$$

In equation (7), α is the TC of threshold voltage, which has a negative value. α is not constant and depends on the transistor dimentions [26]. To compensate temperature effects, expression (6) must be differentiated by temperature and set to zero. Therefore, by substituting (7) in (6), (8) is obtained for temperature compensation as follows:

$$K = \frac{C_{ox.H}}{C_{ox.L}} exp\left(\frac{q}{k_B} \cdot \frac{n_H(-\alpha_1 + \alpha_2 + \alpha_4) - n_L\alpha_3}{n_H n_L}\right) \quad (8)$$

where $K = ((W/L)_4 \cdot (W/L)_2)/((W/L)_3 \cdot (W/L)_1)$. Therefore, temperature compensation can be achieved by choosing the appropriate aspect ratio of transistors. By placing (8) in (6), the temperature compensated output reference voltage will be obtained according to (9).

$$V_{REF} = \frac{n_L(V_{TH.3}(T_0) - \alpha_3 T_0) + n_H(\Delta V_{TH.1} + \Delta\alpha \cdot T_0)}{n_H + n_L} \quad (9)$$

where $\Delta V_{TH.1} = V_{TH.1}(T_0) - V_{TH.2}(T_0) - V_{TH.4}(T_0)$ and $\Delta\alpha = \alpha_2 + \alpha_4 - \alpha_1$.

# 3- LS improvement

In this paper, the LS is improved using a two stage structure and a DIBL compensation block. In the two-stage structure, the supply of the circuit shown in Figure 7 is generated from another circuit with a low LS value. The complete design of the proposed circuit is given in Figure 8, and the voltage reference generator given in [12] is used as the first stage. Using the two-stage structure, the overall LS can be obtained by multiplying the LS values of each stage [14]:

$$LS_{Total} = LS_1 \times LS_2 \quad (10)$$

where $LS_1$ is the line sensitivity of the first stage and $LS_2$ is the line sensitivity of the second stage. The first and second stages are referred to as LS and TC correctors, respectively. The TC correction stage requires a minimum voltage of $4V_T+V_{PTAT}$ for correct operation. Because of this, a short channel length is chosen for $M_{15}$ to supply the minimum voltage required in the second stage. The short length of this current source transistor in the LS correction stage increases the DIBL effect. It is not possible to use the structure of [22] to compensate the DIBL effect, as it requires access to a ratio of $I_{D15}$ current. The solution presented here aims to provide DIBL compensation using transistors $M_{11}$, $M_{12}$, and $M_{13}$.

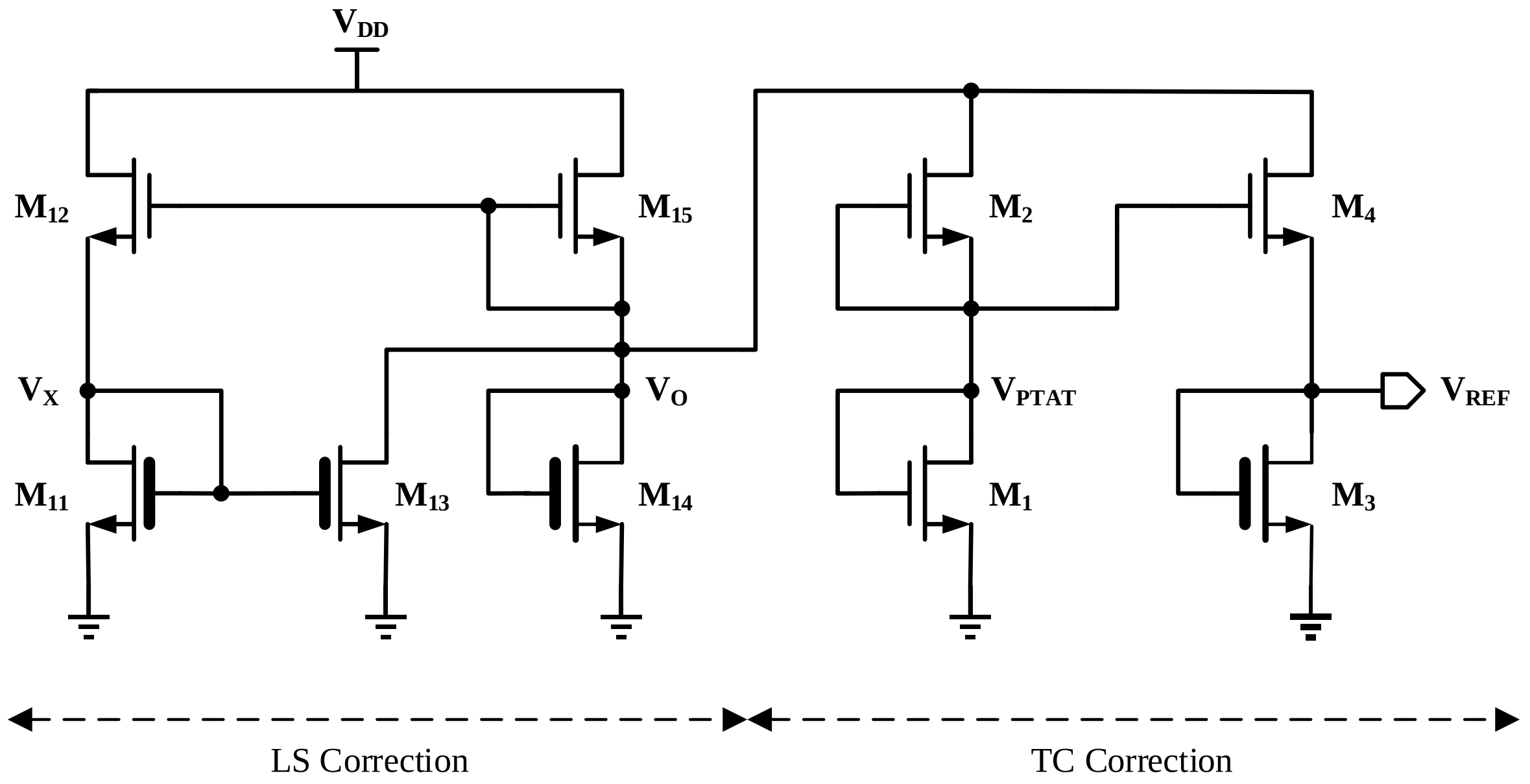

Figure 8- Complete architecture of the proposed voltage reference

To extract the total LS, it is necessary to obtain the LS of each stage separately according to (10). The LS of the first stage is obtained using its output voltage equation. By considering the equal currents on transistors $M_{11}$ and $M_{12}$, the voltage $V_X$ is described by

$$V_X = \frac{n_L V_{TH.11} - n_H V_{TH.12} + n_H n_L V_T \ln(K_{R1}) + n_H V_o}{n_H + n_L} \tag{11}$$

where $K_{R1} = (C_{ox.L} \cdot (W/L)_{12})/(C_{ox.H} \cdot (W/L)_{11})$. $I_{D,13}$ can be achieved using (11) as follows:

$$I_{D.13} = \mu_n C_{OX.H} (\frac{W}{L})_{13} V_T^2 \exp\left(\frac{\Delta V_{TH.2} + n_H n_L V_T\, ln(K_{R1}) + n_H V_o}{n_H(n_H + n_L)V_T}\right) \tag{12}$$

where $\Delta V_{TH.2} = n_L V_{TH.11} - n_H V_{TH.12} - (n_H + n_L)V_{TH.13}$. On the other hand, the current $I_{D15}$ is obtained as

$$I_{D.15} = \mu_n C_{OX.L} (\frac{W}{L})_{15} V_T^2 \exp\left(\frac{-V_{TH.15}}{n_L V_T}\right) \tag{13}$$

$I_O$ is found by subtracting $I_{D,13}$ from $I_{D,15}$. Subsequently, $V_O$ is obtained by substituting $I_O$ in the gate-source voltage expression of $M_{14}$.

$$V_O = V_{TH.14} + n_H V_T \ln\left(K_{R2}\, exp\left(\frac{-V_{TH.15}}{n_L V_T}\right)\left[1 - K_{R3}\, exp\left(\frac{\Delta V_{TH.3} + n_H n_L^2 V_T\, ln(K_{R1}) + n_H n_L V_o}{n_H n_L (n_H + n_L) V_T}\right)\right]\right) \tag{14}$$

where $\Delta V_{TH.3} = n_L^2 V_{TH.11} - n_H n_L V_{TH.12} - n_L(n_H + n_L)V_{TH.13} + n_H(n_H + n_L)V_{TH.15}$, $K_{R2} = \left(C_{ox.L}\,(W/L)_{15}\right)/\left(C_{ox.H}\,(W/L)_{14}\right)$ and $K_{R3} = \left(C_{ox.H}\,(W/L)_{13}\right)/\left(C_{ox.L}\,(W/L)_{15}\right)$.

To further simplify (14) the expression inside the bracket should be simplified. For this purpose, the linearization given in (15) will be used in which $X = K_{R3}\, exp\left(\frac{\Delta V_{TH.3} + n_H n_L^2 V_T\, ln(K_{R1}) + n_H n_L V_o}{n_H n_L (n_H + n_L) V_T}\right)$.

$$\ln(1 - X) = C_1X + C_2 \tag{15}$$

Using the technical parameters, the *X* value range is obtained in the range of [0.57,0.85]. Figure 9 shows the plot of *ln(1-X)* and its linearized approximate. According to Figure 9, the linearization is a good approximation of *ln(1-X)* with a correlation coefficient, $R^2$, of 0.98. In the linearization coefficients, $C_1$ and $C_2$ are -3.64 and 0.98, respectively.

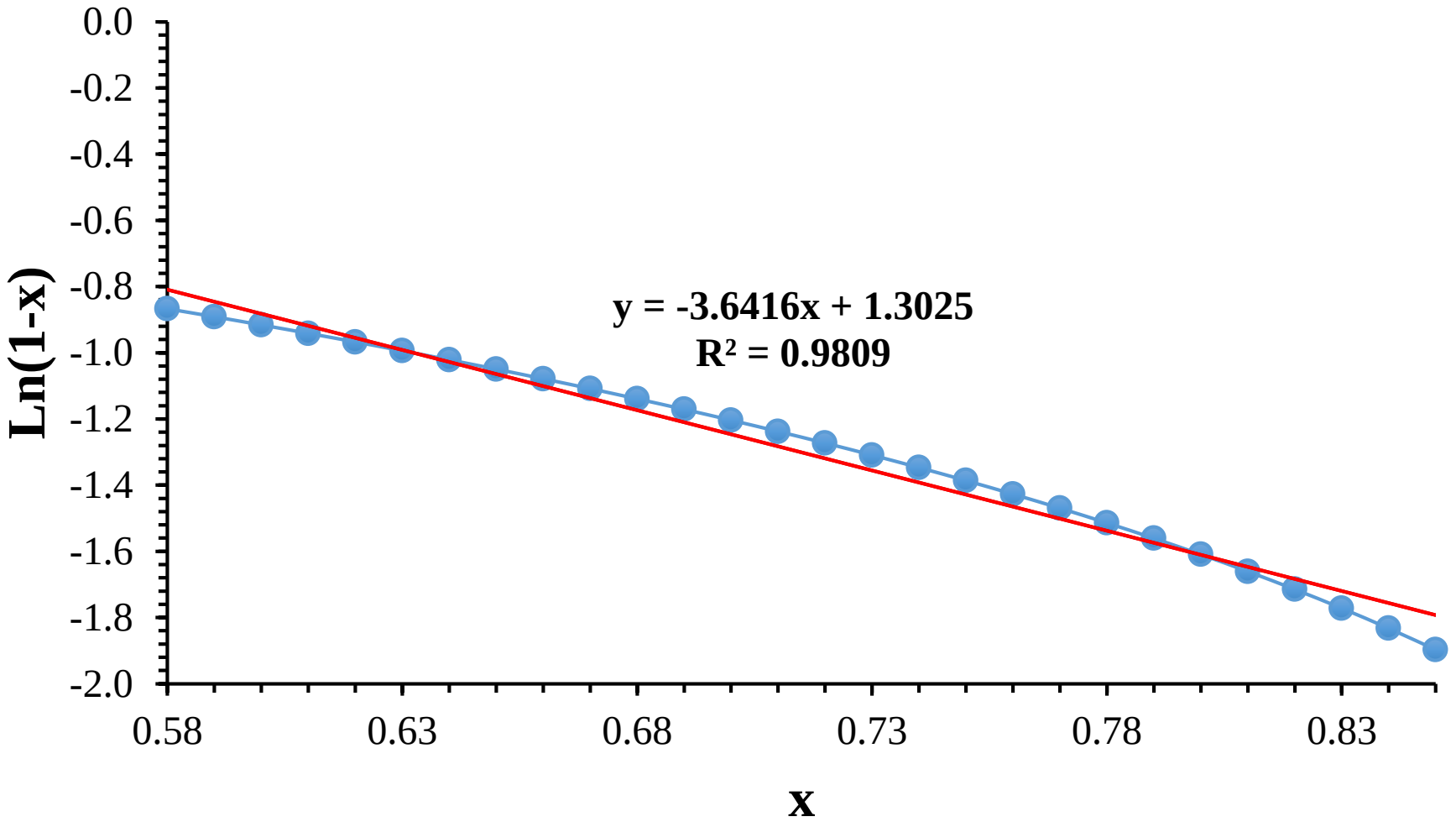

Figure 9- linearization in expression (12)

Substituting (15) in (14) and simplifying the result, (16) is obtained as follows:

$$V_O = \frac{\Delta V_{TH.4} + n_H V_T (C_3 n_L^2 \ln K_{R1} + C_2)}{1 - C_3 n_H n_L} \tag{16}$$

where $C_3 = C_1/(n_L(n_L + n_H))$, $\Delta V_{TH.4} = V_{TH.14} - N_R V_{TH.15} + C_3 \Delta V_{TH.3}$ and $N_R = n_H/n_L$. The main dependence of the output voltage on the supply voltage is through the DIBL effect on the threshold voltage in the sub-threshold region. This effect can be expressed as follows [22,27]:

$$V_{TH} = V_{TH.0} - \lambda_D V_{DS} \tag{17}$$

where $V_{TH,0}$ is the threshold voltage with $V_{DS}=0$ and $\lambda_D$ is the DIBL factor. Transistors $M_{12}$ and $M_{15}$ suffer from substantial $V_{DS}$ changes and therefore, in order to compensate the DIBL effect, the terms related to the DIBL factor of these two transistors should be removed from (16). Thus according to (17), the DIBL effect can be compensated using (18) as follows:

$$\lambda_{D.12} = C_4(1 + N_R)\lambda_{D.15} \tag{18}$$

where $C_4=(C_1-1)/C_1$. In this condition in the approximated equations, $LS_1$ will be equal to zero however in practice due to second-order effects a non-zero $LS_1$ will eventually be achieved.

In the following, LS of the TC correction stage is obtained. For this purpose, (9) is differentiated with respect to $V_O$.

$$\frac{\sigma V_{REF}}{\sigma V_O} = \frac{n_H}{n_H + n_L} \cdot (\lambda_{D.2} + \lambda_{D.4}) \tag{19}$$

Therefore, the LS of the second stage is obtained as follows:

$$LS_2 = \frac{n_H}{(n_H + n_L)V_{REF}} \cdot (\lambda_{D.2} + \lambda_{D.4}) \times 100 \tag{20}$$

Considering (20) the LS of the second stage alone is not suitable and the first stage can help achieve a desirable LS value. In order to show the effect of the methods used in improving the LS, the output reference voltage without using the LS correction stage, using the two-stage structure (without DIBL compensator) and finally using the LS correction stage is shown in Figure 10. The LS of these structures in the typical corner are 0.542%/V, 0.026%/V, and 0.013%/V, respectively, which indicates the effectiveness of the methods used to improve LS.

It should be noted that in order to achieve an accurate expression for temperature compensation, the effect of $V_O$ changes with temperature in the first stage on the output reference voltage should be investigated. The effect of these changes on the reference voltage can be expressed by (21) [14].

$$\left(\frac{\partial V_{REF}}{\partial V_O} |T = cte \times \frac{\partial V_O}{\partial T}\right) / V_{REF} = LS_2 \times TC_{LS_Comp} \times V_{O-avg} \approx 0 \tag{21}$$

Due to the very small effect of the first stage TC on the output reference voltage, it's effect is neglected here.

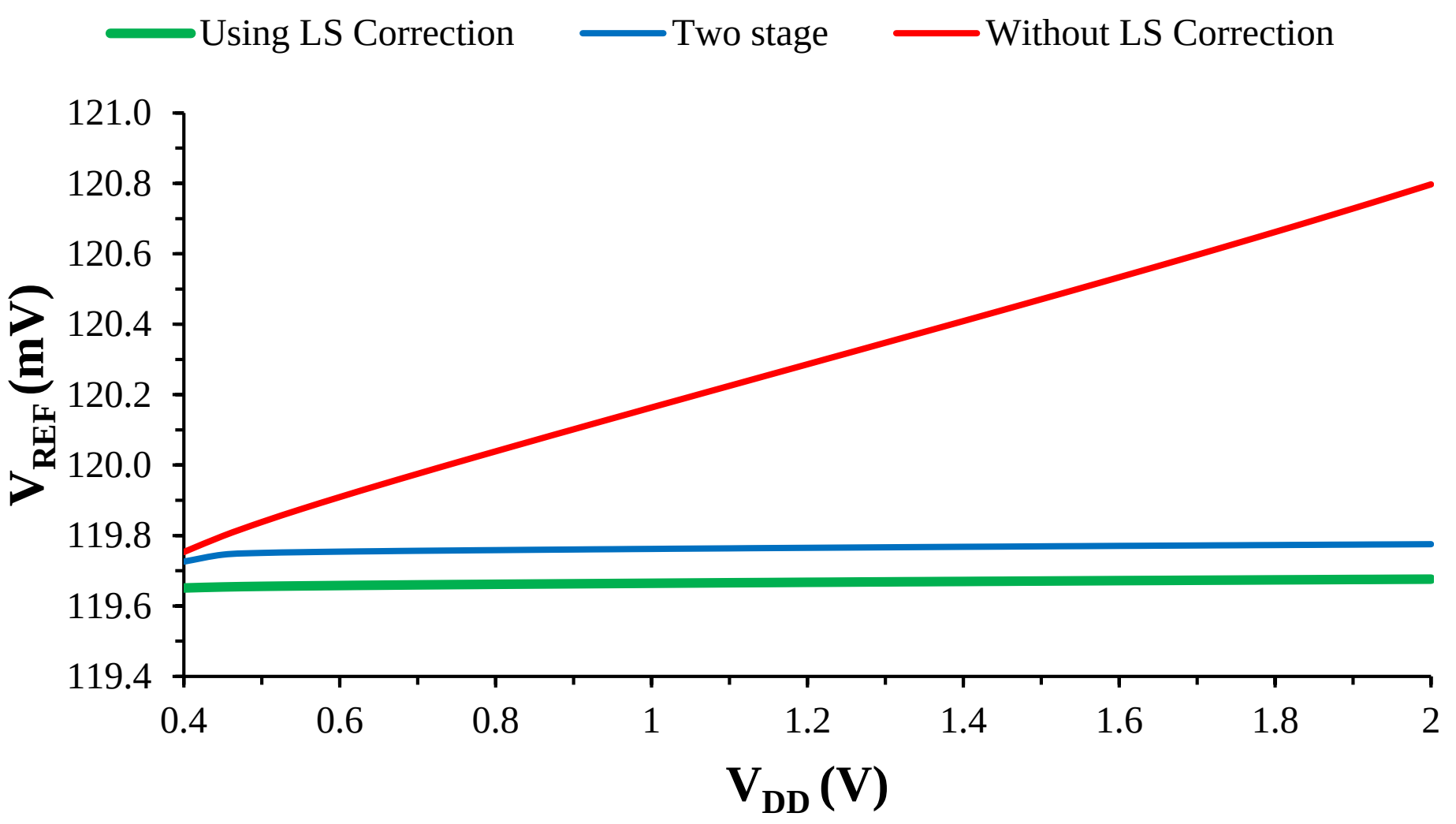

Figure 10– Investigation of methods used to improve line sensitivity

## 4- PSRR analysis

In the "LS improvement" section, power supply voltage fluctuations were examined using the large-signal model (DC state). However, the parasitic capacitance effect of MOSFETs can cause the supply potential noise to leak out at high frequencies. Therefore, it is important to examine the effect of power supply frequency components on the reference voltage in the frequency domain. Figure 11 shows the simplified circuit of the proposed circuit assuming that the TC correction stage loading is neglected and $V_{PTAT}$ is assumed constant. In the circuit model, $c_1$ and $c_3$ are capacitances between $v_{REF}$ and $v_O$ to ground, respectively, while $c_2$ and $c_4$ are the capacitances between $v_{REF}$ and $v_O$ to $v_{DD}$. $r_{o2}$ and $r_{o4}$ are the output resistances of $M_2$ and $M_4$, respectively. In addition, $g_{m1}$, $g_{m3}$, and $g_{m14}$ are the trans-conductance of $M_1$, $M_3$, and $M_{14}$, respectively. The DIBL compensation circuit cancels the effect of $r_{o15}$, so it is omitted in Figure 11. Thus the PSRR of the proposed circuit is given as

$$\frac{\partial V_{REF}}{\partial V_{DD}} = \frac{\partial V_{REF}}{\partial V_O} \times \frac{\partial V_O}{\partial V_{DD}} \approx \frac{2r_{o4}c_2 s + 1}{2r_{o4}c_2 s + 2r_{o4} + 1} \times \frac{c_4 s}{g_{m.14} + (c_3 + c_4)s} \quad (22)$$

Equation (22) can be rewritten as follows

$$\frac{\partial V_{REF}}{\partial V_{DD}} \approx \frac{c_4}{g_{m.14}(2r_{o4} + 1)} \cdot \frac{s(1 + s/z_1)}{(1 + s/p_1)(1 + s/p_2)} \quad (23)$$

where $z_1=1/2r_{o4}c_2$, $p_1=(2r_{o4}+1)/2r_{o4}c_2$ and $p_2=g_{m,14}/(c_3+c_4)$. According to (23), there are two poles and two zeros. One zero is near the origin, while the other ($z_1$) is further away. The poles are expected to be between the two zeros.

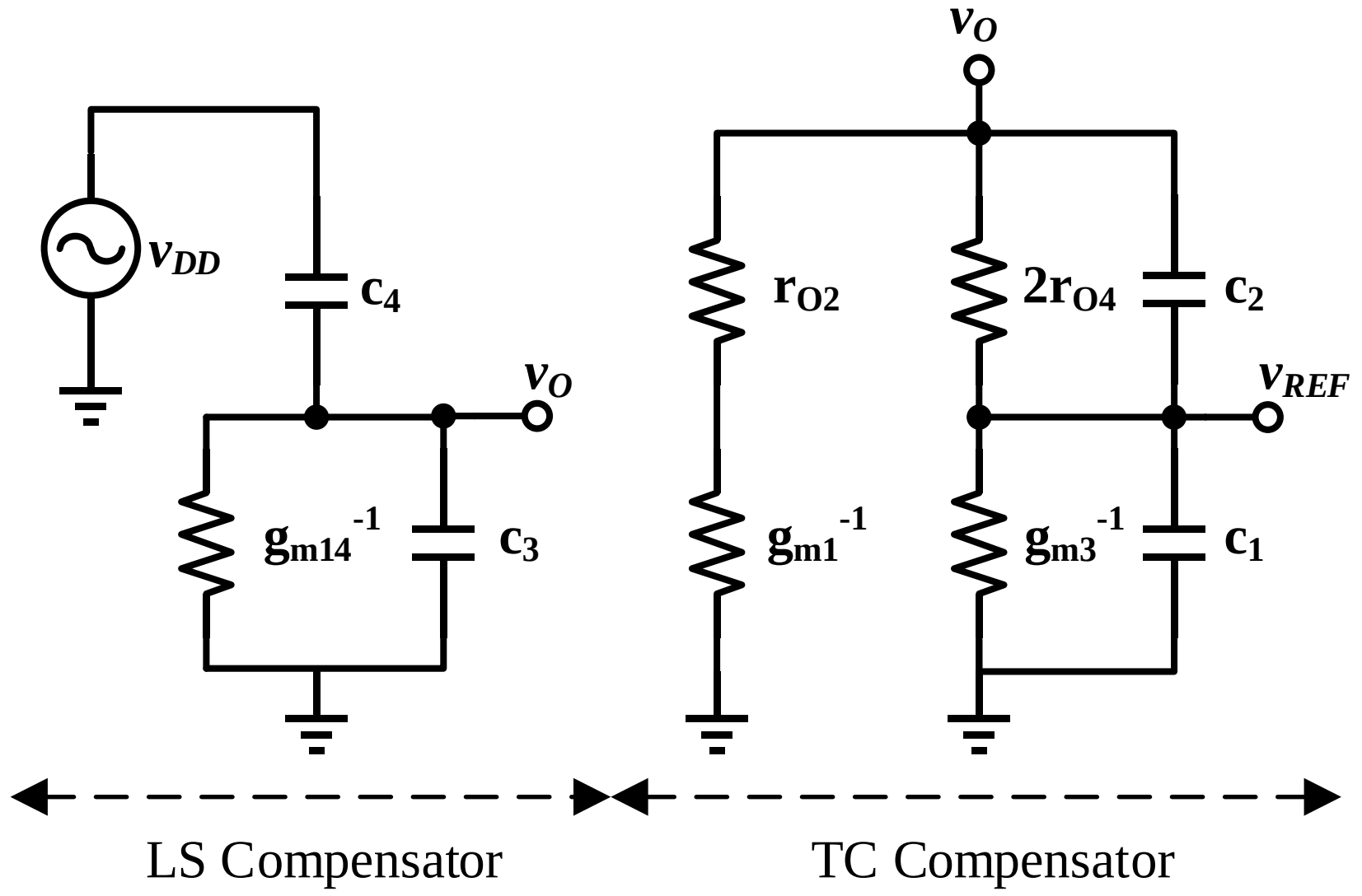

Figure 11- AC equivalent of the proposed circuit

The PSRR of the proposed circuit with and without the LS correction stage at 1V supply voltage is plotted in Figure 12 for frequencies from 1 Hz to 10 GHz with the temperature adjusted at 25 ºC. The minimum PSRR for both are at 1 Hz, the value of which, with and without the use of the LS correction stage are -97 dB and -63.6 dB respectively, while the maximum values are at 190 Hz and 53 Hz, respectively, and equal to -71 dB and -45 dB. At the upper frequency, the PSSRs decrease to about -78 dB and -62.4 dB, respectively. As expected, in the proposed structure, the first zero near the origin increases the PSRR; however, the next two poles lead to a decline. The high frequency zero eventually prevents the trend decline and results in a flat response.

Considering the obtained result, it can be observed that the use of LS correction stage significantly improves PSRR.

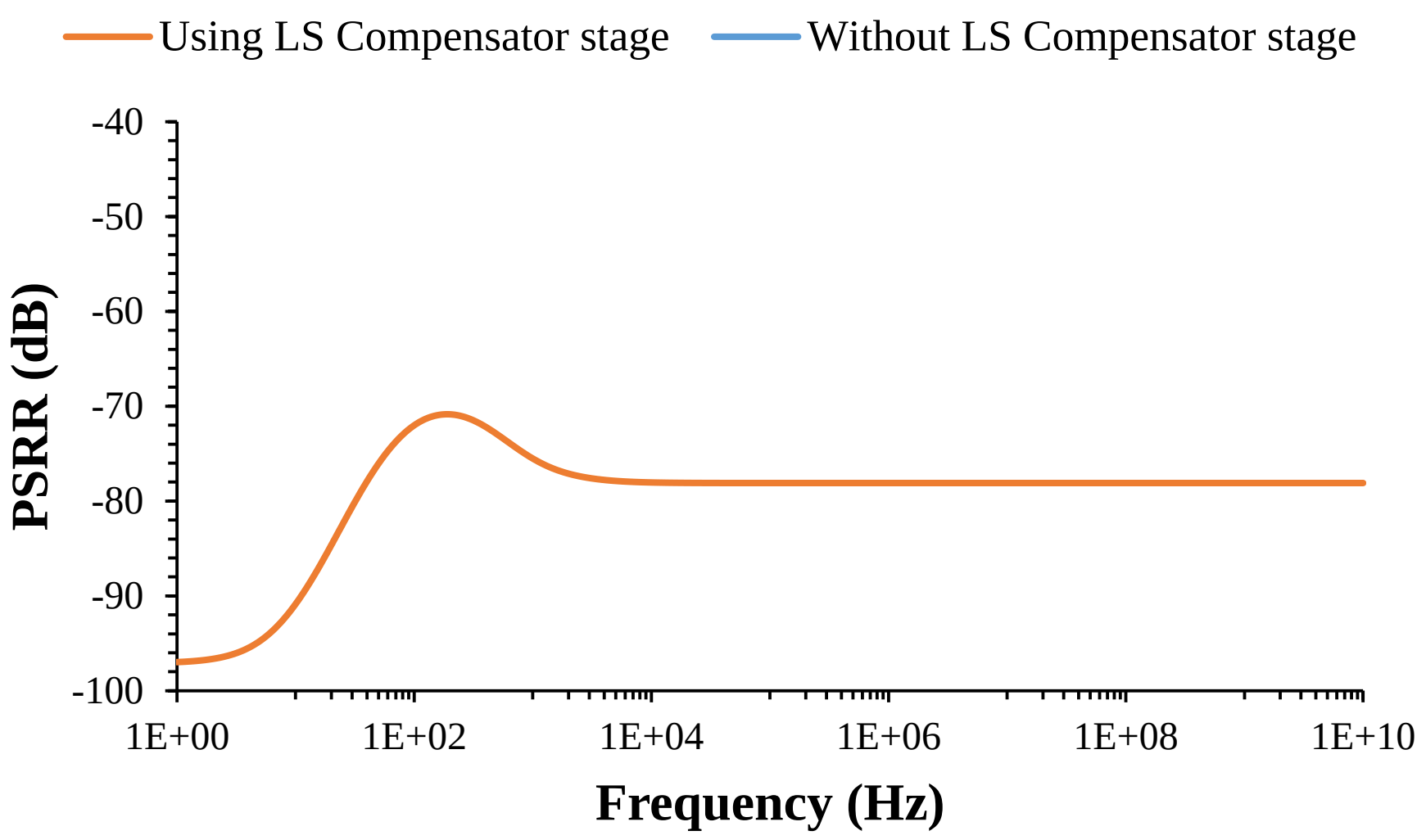

Figure 12- PSRR of the proposed design without using the LS correction stage and using it in the frequency range of 1 Hz to 10 GHz

# 4- Trimming circuit

Various characteristics of MOSFETs vary at different corners. One of them is the threshold voltage which has a key role in the output voltage reference. A trim circuit can improve TC at different corners, but it can also increase complexity and area. Various solutions such as current trim [28] and voltage dividers [29] can be used to improve the accuracy of the output reference voltage. In this paper, although it has been verified in the simulations that an acceptable TC is achieved under different corners and mismatch conditions however a simple trimming method can be used for scenarios that need a better TC and wider temperature operation range. The trimming circuit is shown in Figure 13 which is similar to the trimming solutions presented in [23] and [22]. By adjusting the effective aspect ratio of transistor $M_3$ in different corners, it is possible to reach the required optimum K in different corners. The optimal K can be calculated from (8). Similar to [23] three binary codes are used to tune the size of $M_3$. SS and FF corners determine the maximum and minimum $M_3$ transistor size for TC compensation. So, $M_{3\text{-}T2}$ to $M_{3\text{-}T0}$ are chosen such that compensation can be covered in these extreme corners. Figure 14 plots $V_{REF}$ versus temperature from -10 °C to 85 °C while in each case the suitable trim configuration is chosen. TC values for TT, FF, FS, SF, and SS are 8.84 ppm/ºC, 18.25 ppm/ºC, 14.48 ppm/ºC, 27.39 ppm/ºC, and 35.12 ppm/ºC, respectively.

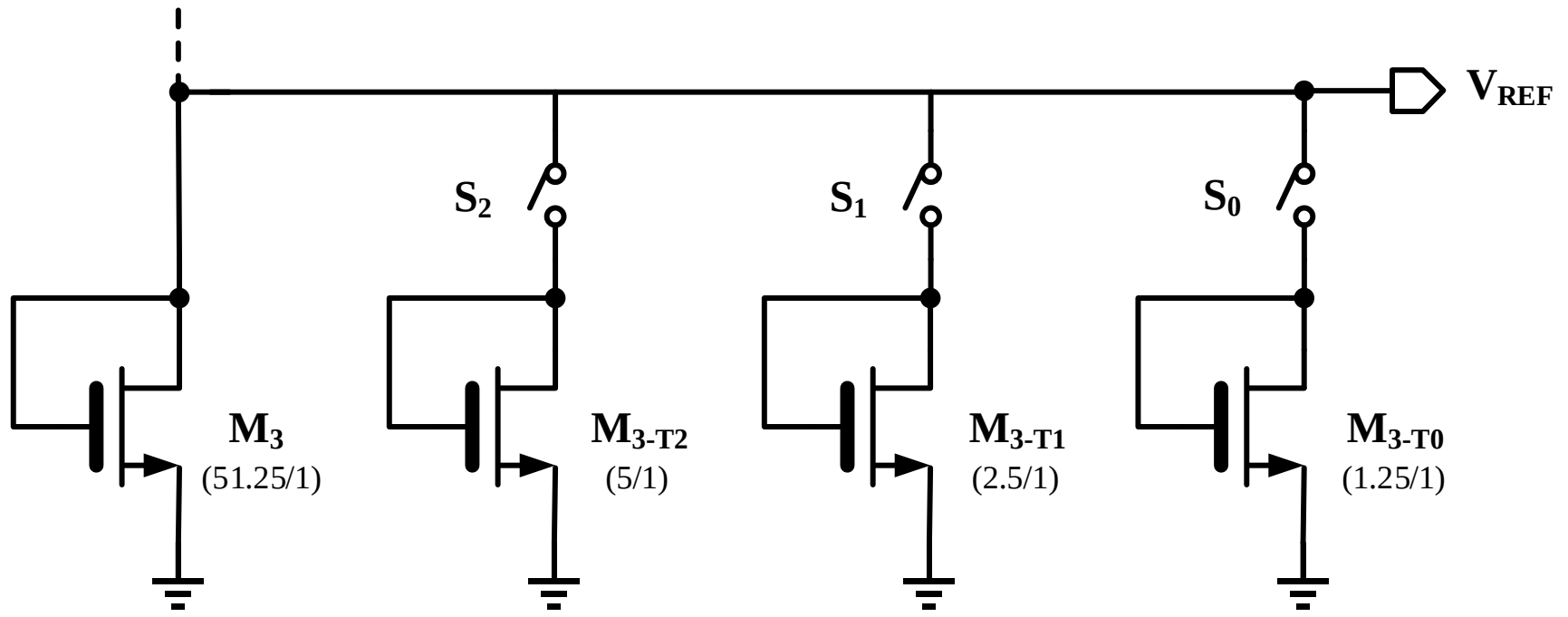

Figure 13- Trim circuit

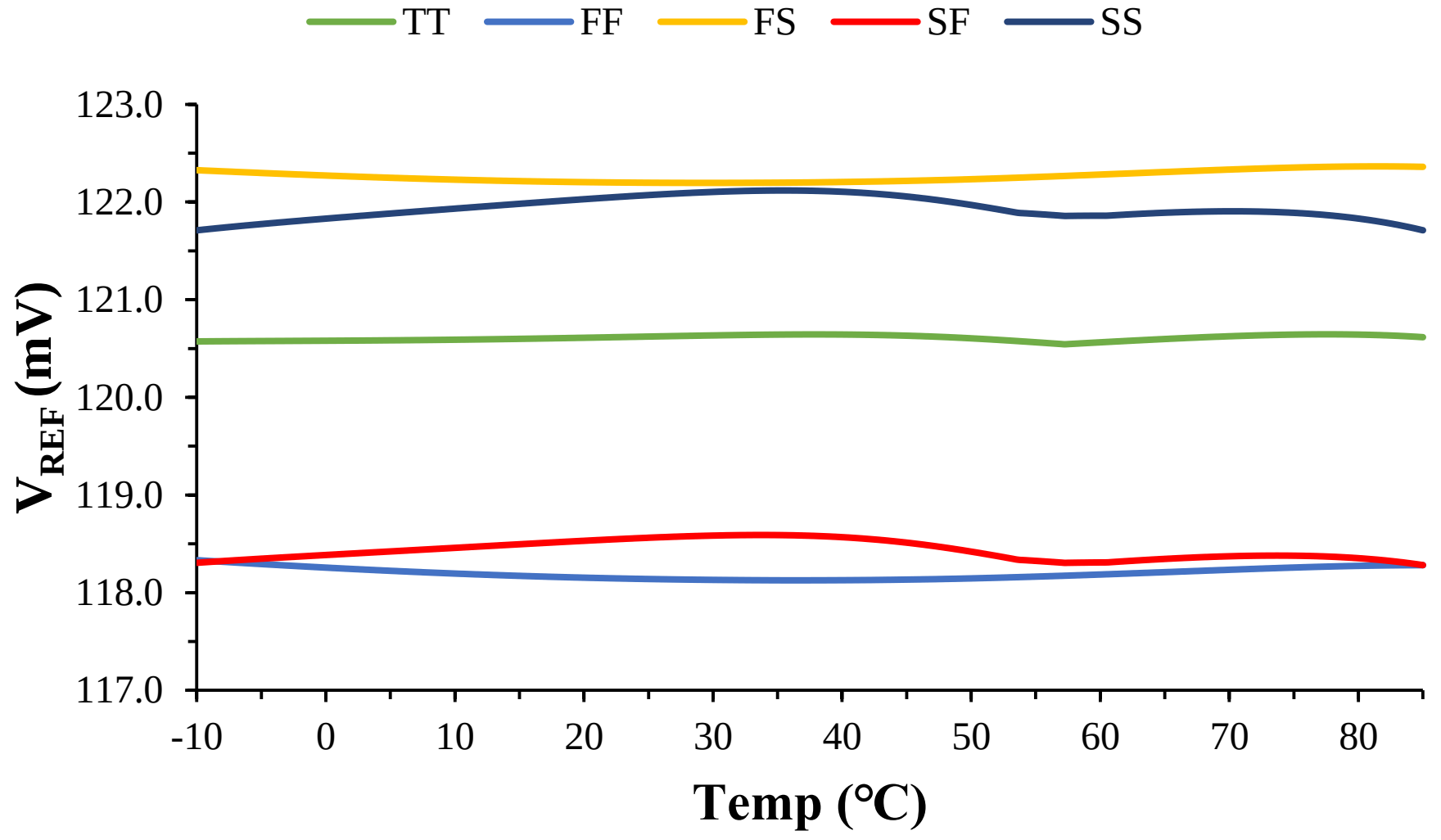

Figure 14- $V_{REF}$ versus temperature after the trim at different corners

# 5- Optimal aspect ratio for temperature compensation

Equation (8) provides the appropriate dimension ratio for temperature compensation. Figure 15 shows the TC changes against the K ratio. In this figure, using simulation, an optimal value of TC equal to K=3.87 is obtained, which is very close to the value K=3.83 obtained by theoretical elaborations obtained in (8). Therefore, (8) gives a good prediction on the optimal point for temperature compensation.

The sensitivity of the reference voltage to changes in the power supply causes a change in TC. Due to the LS correction stage, small TC changes relative to its optimal values are to be expected. Figure 16 shows the TC changes against the supply voltage with and without the LS correction stage. Thus the use of an LS correction stage, in addition to improving LS, brings stability to the obtained TC against supply voltage changes.

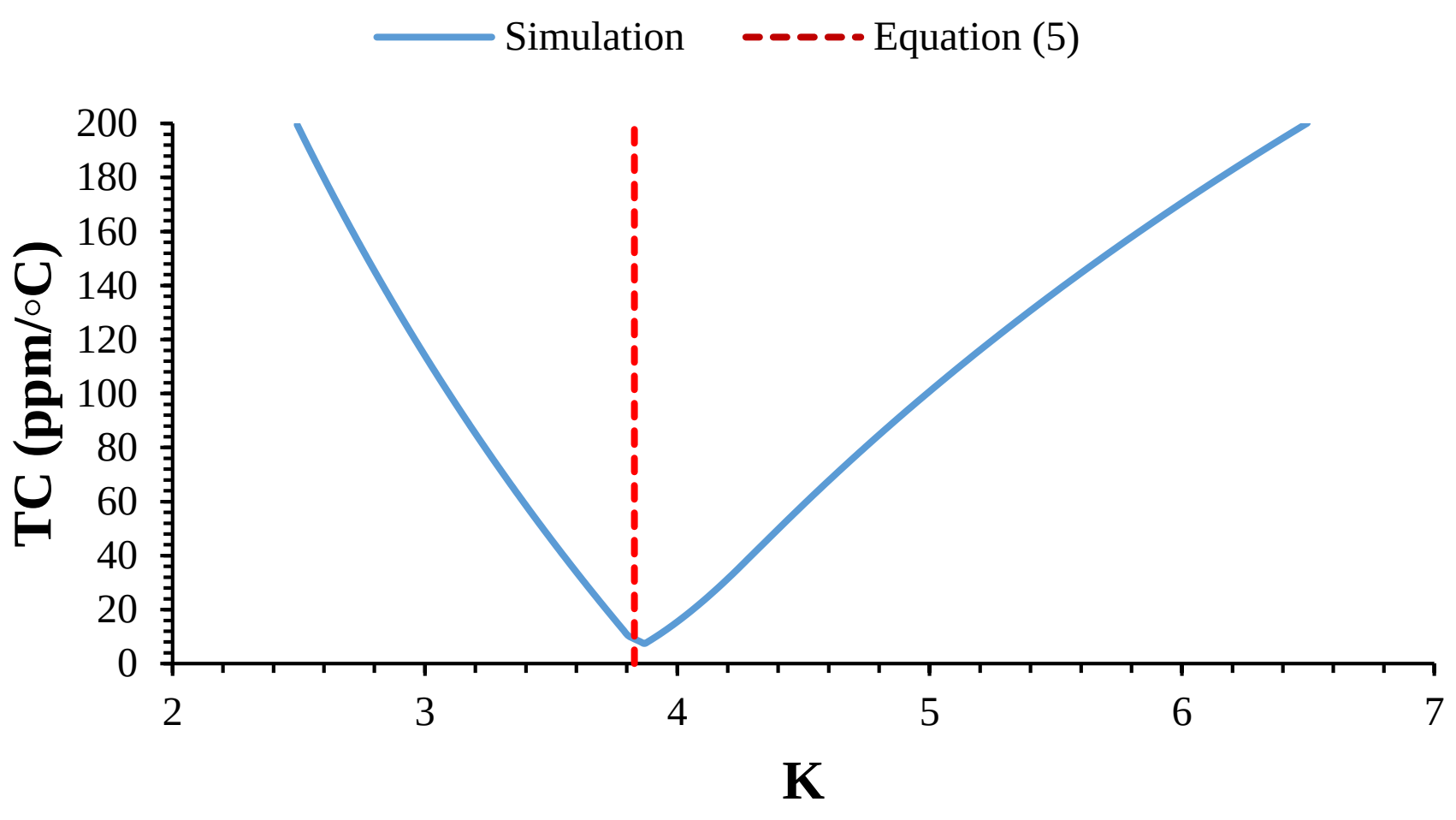

Figure 15- TC changes as a function of K ratio to compare the optimal K in simulation versus Equation (8)

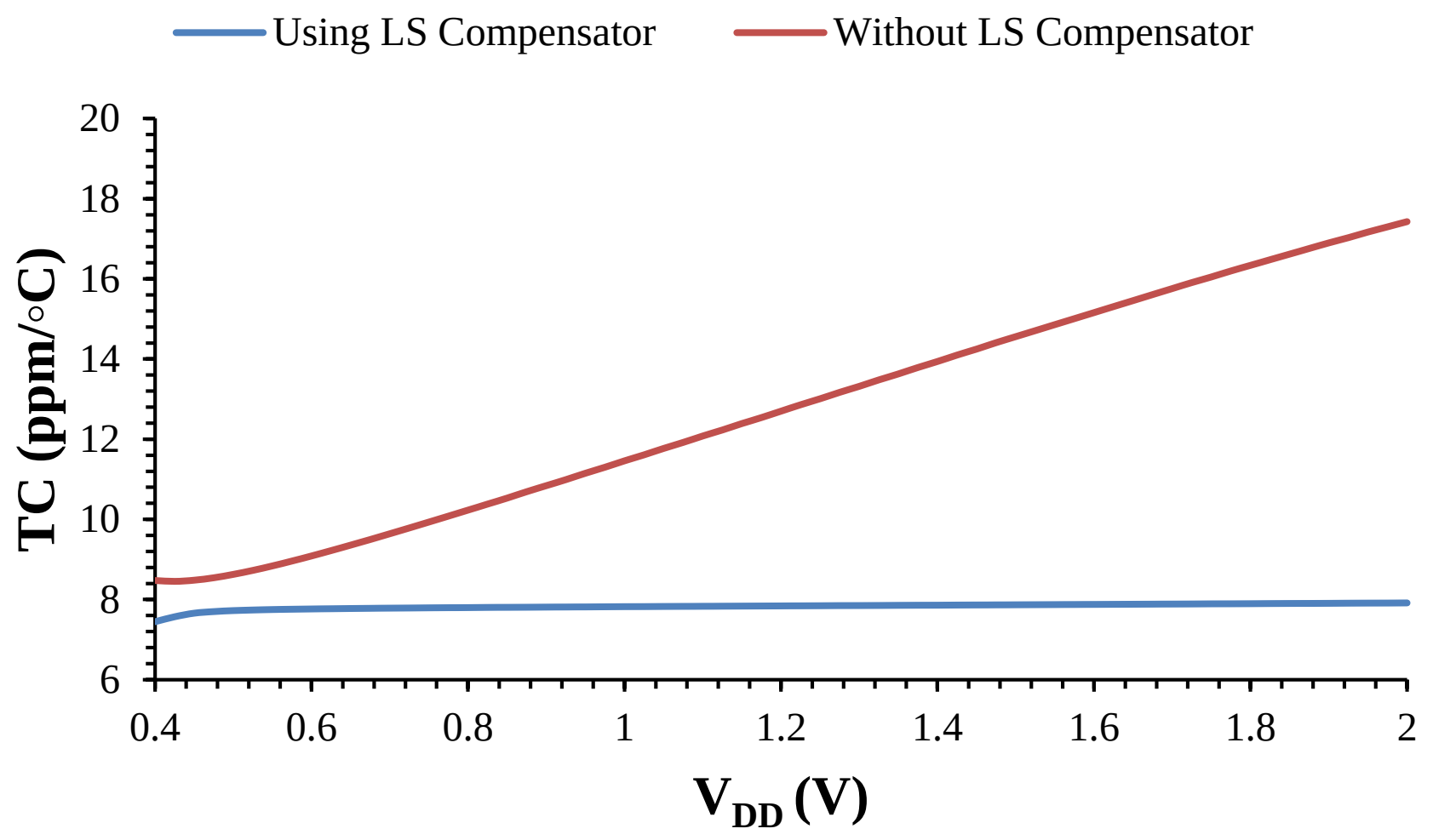

Figure 16- TC changes with and without the LS correction stage

# 6- Design dimensions and layout

For the final design, the MOSFET dimensions of the second stage are chosen based on the optimal aspect ratio obtained in the previous section for temperature compensation. Long channel devices are chosen in the second stage to achieve a suitable LS. However since the first stage has the main contribution for correcting LS, the DIBL compensation method is used in the first stage. In addition, at worst case conditions, the minimum output potential for the first stage is chosen based on the minimum supply potential of $4V_T+V_{PTAT}$ for the second stage which results in near $V_{DS}$ independent subthreshold operation of the second stage MOSFETs. Eventually, the MOSFET dimensions in the first stage are chosen based on three factors. Both channel lengths and widths in the first stage are chosen to 1- meet the target area, 2- produce the required output potential for the second stage, and 3- to generate the required DIBL compensation over the voltage reference operation range. Long channel devices in the first stage can lower the LS however as shown in Figure 17, with the DIBL compensation better LS can be achieved since with the DIBL compensation, the response contains an extremum which can result in a near flat response over a specific voltage supply range. Also using a long channel device for $M_{15}$ will result in a very large channel width for this MOSFET, increasing the area and also parasitic capacitance effects and hence degrading PSRR.

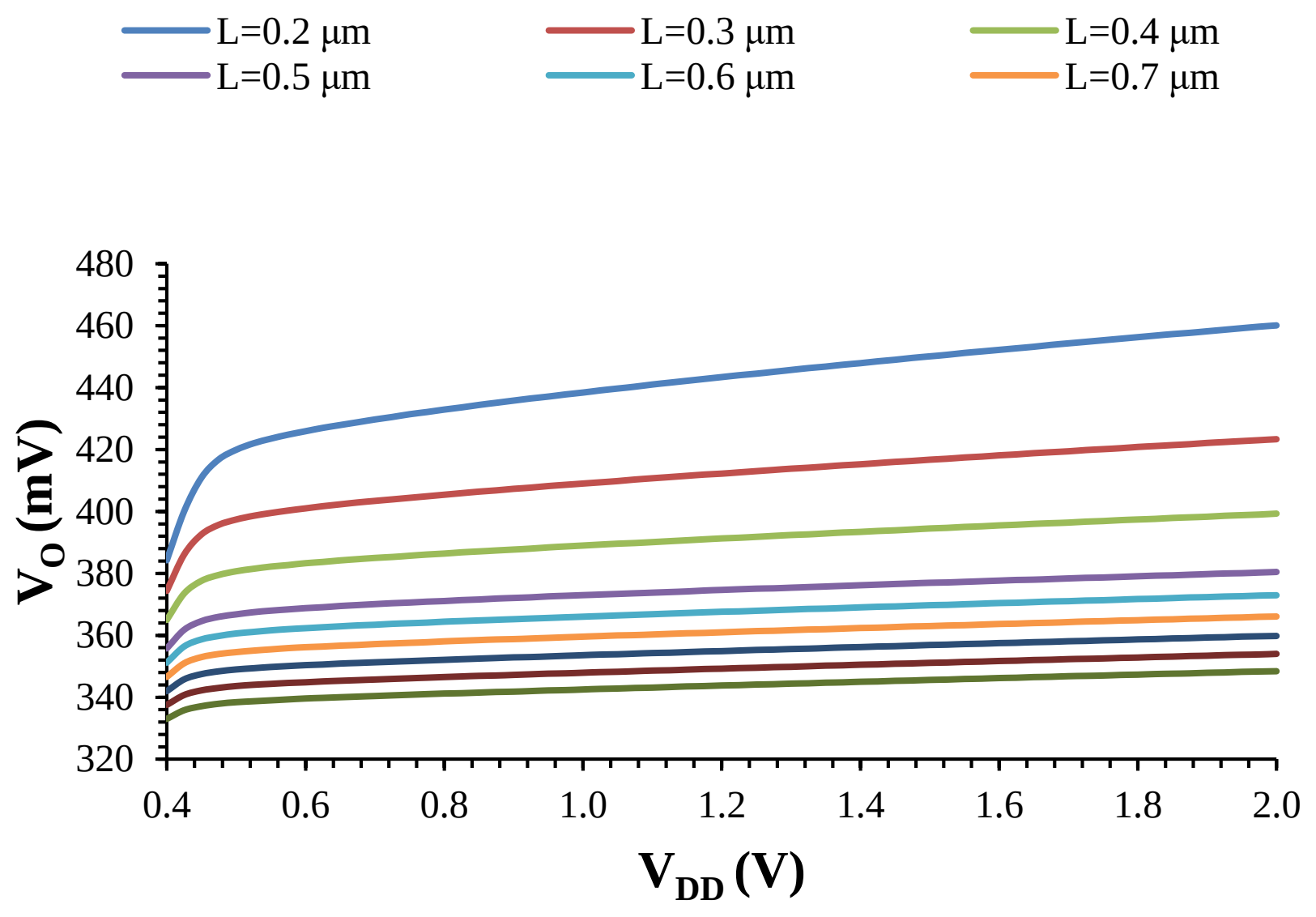

(a)

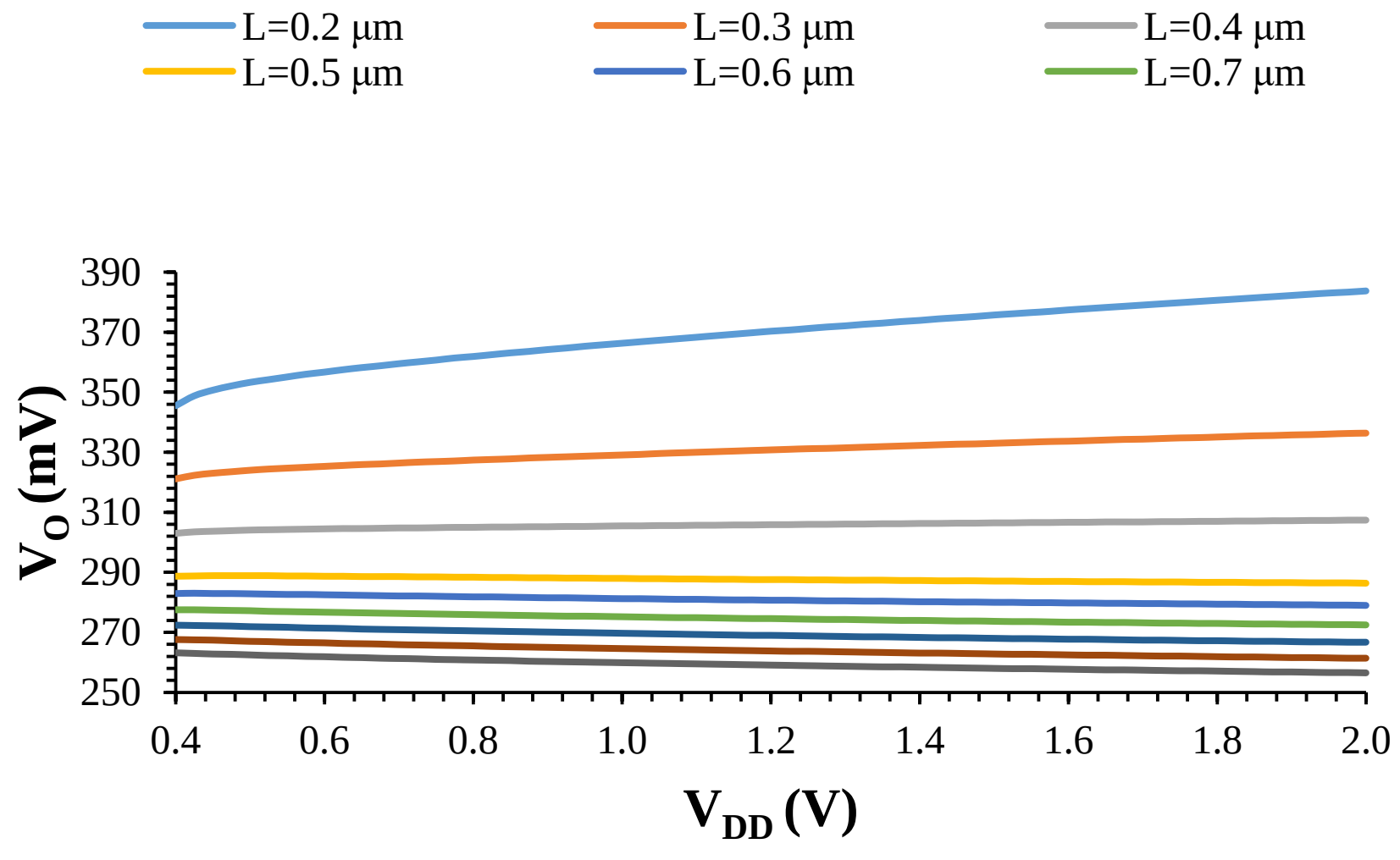

(b)

Figure 17- $V_O$ versus $V_{DD}$ with different $M_{15}$ lengths in (a) two stage (b) proposed structure

The transistor sizes of the proposed circuit are reported in Table 1. The effects of variation during fabrication are also investigated with a post-layout simulation and Monte Carlo analysis with 1000 runs. Figure 18 shows the layout of the proposed architecture. The area of the proposed circuit is about 2183 μm$^2$.

Table 1- The size and type of transistors in the proposed voltage reference

| **Transistor** | **Type** | **Size (W/L)** |
|---|---|---|
| **$M_1$** | Thin oxide- NMOS | 1 μm/20 μm |
| **$M_2$** | Thin oxide- NMOS | 38.1 μm/5 μm |
| **$M_3$** | Thick oxide- NMOS | 57.5 μm/1 μm |
| **$M_4$** | Thin oxide- NMOS | 29.2 μm/20 μm |
| **$M_{11}$** | Thick oxide- NMOS | 1 μm/1 μm |
| **$M_{12}$** | Thin oxide- NMOS | 20 μm/0.18 μm |
| **$M_{13}$** | Thick oxide- NMOS | 4 μm/0.5 μm |
| **$M_{14}$** | Thick oxide- NMOS | 0.22 μm/20 μm |
| **$M_{15}$** | Thin oxide- NMOS | (60 μm/0.3 μm)×3 |

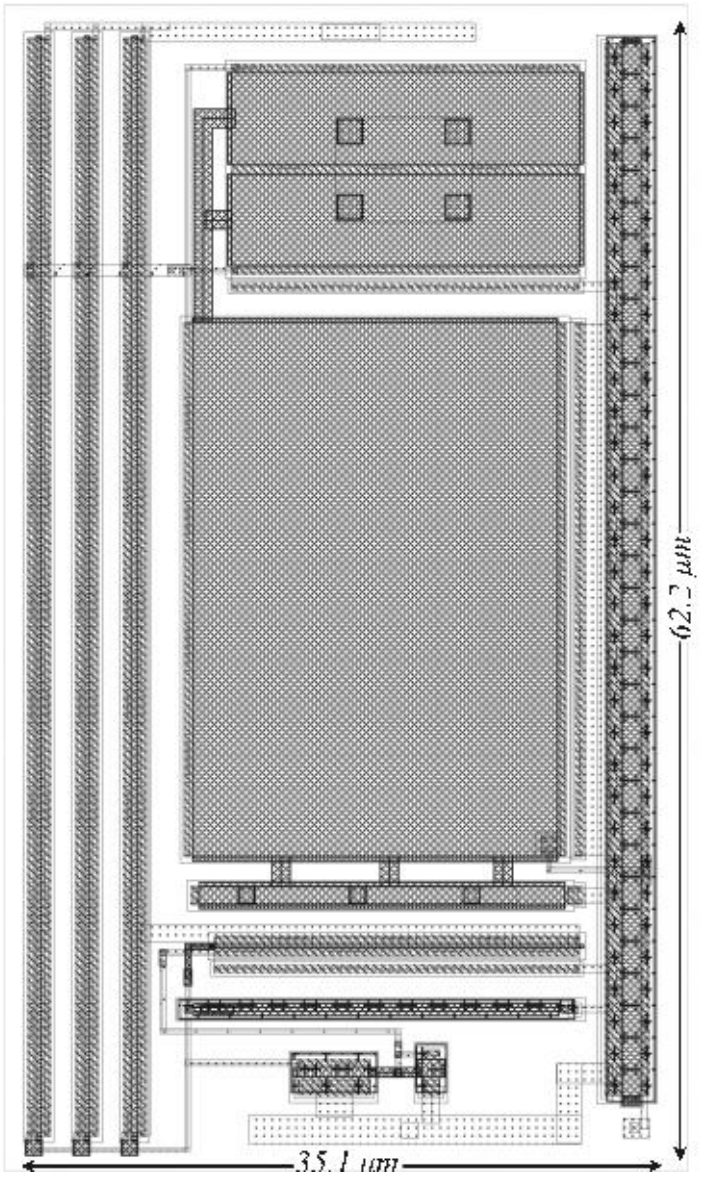

Figure 18- The layout of the proposed voltage reference

# 7- Evaluation results

The voltage reference circuit proposed in this paper is simulated using a 0.18 μm standard CMOS technology. Since self-bias circuits using feedback paths are not used in the proposed design, a start-up circuit will not be required, reducing the occupied area.

The changes of the output reference voltage with respect to the supply in different corners are shown in Figure 19. The minimum supply voltage in this design is 0.4 V and the obtained LS in the TT, FF, SS, FS, and SF corners at 25 °C are 130.24 ppm/V, 164.72 ppm/V, 131.77 ppm/V, 164.48 ppm/V, and 133.3 ppm/V, respectively.

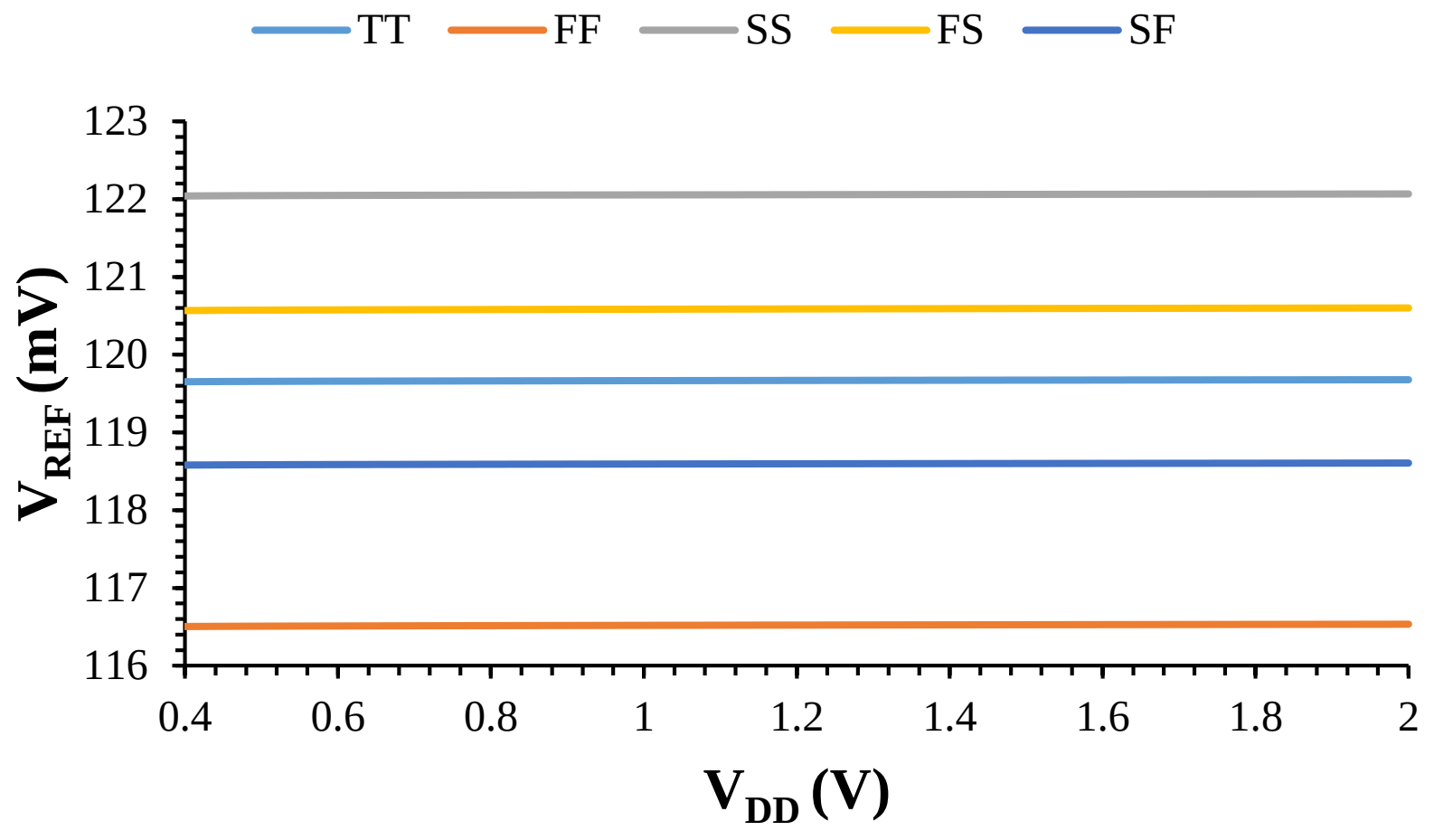

Figure 19- Reference voltage changes against supply voltage at 25 ° C in different corners

The changes of the output voltage with temperature is evaluated in different corners at a supply potential of 0.4 V. Figure 20 shows the output reference voltage variations in the temperature range of 0 °C to 80 °C at the TT, FF, SS, FS, and SF corners. The TCs in these corners are 7.45 ppm/°C, 44.72 ppm/°C, 40.14 ppm/°C, 35.13 ppm/°C, and 35.08 ppm/°C respectively. The output reference voltage according to (6) is substantially influenced by the threshold voltages difference and since the threshold voltage is a process-dependent parameter, it leads to a small shift of the reference voltage in different corners and also increases the TC. These changes can be alleviated by using the trimming procedure presented in this article. Due to the increased complexity and larger area, it is not included in the layout design of this paper.

The changes in power consumption versus temperature is also investigated for supply voltages of 0.4 V, 1.2 V, and 2 V in Figure 21. The minimum power consumption in 0.4 V supply and 0 ℃ is 3.25 pW, and its maximum is 2.84 nW in 2 V and 80 ℃.

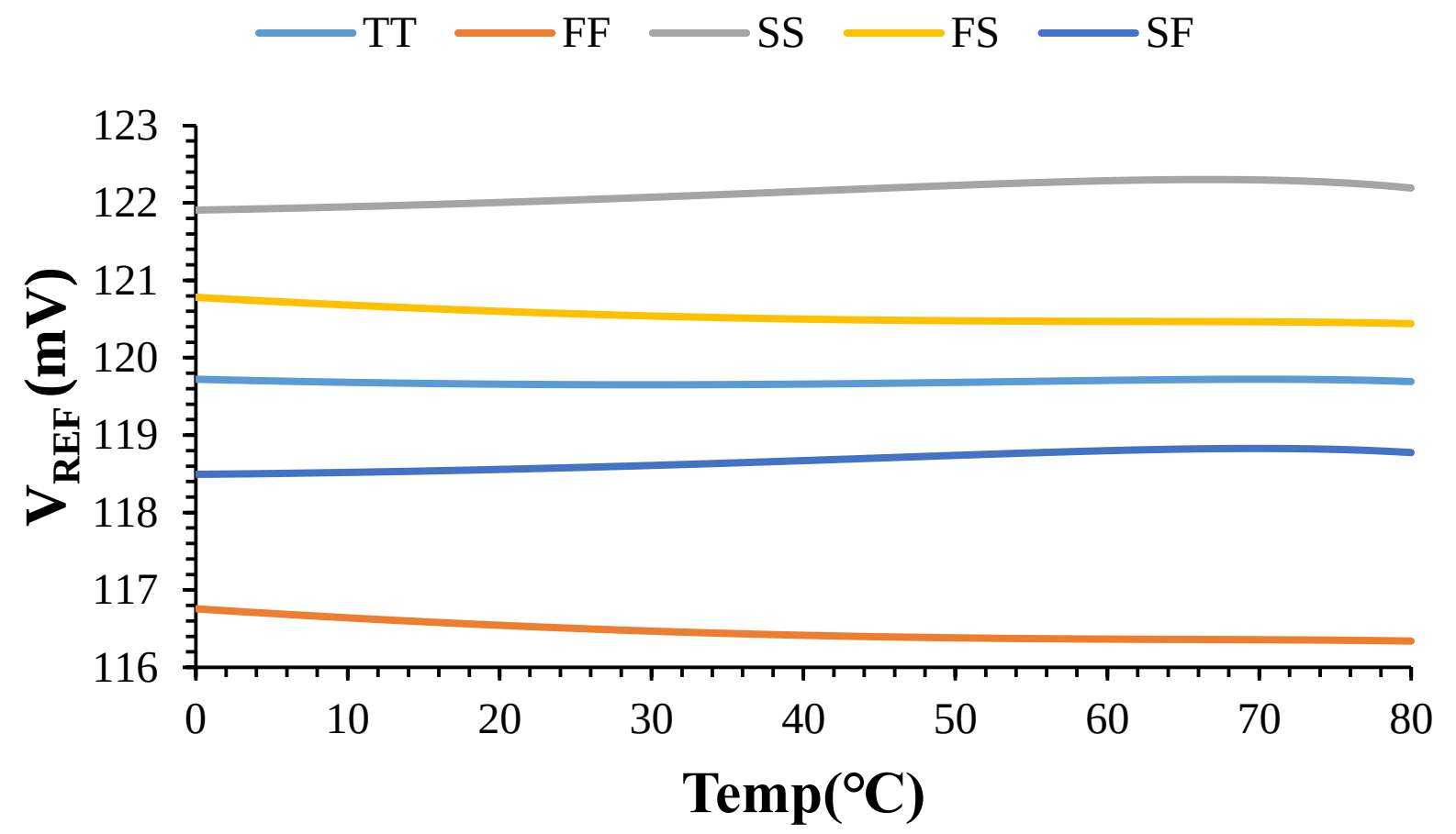

Figure 20- Output reference voltage changes against temperature in different corners

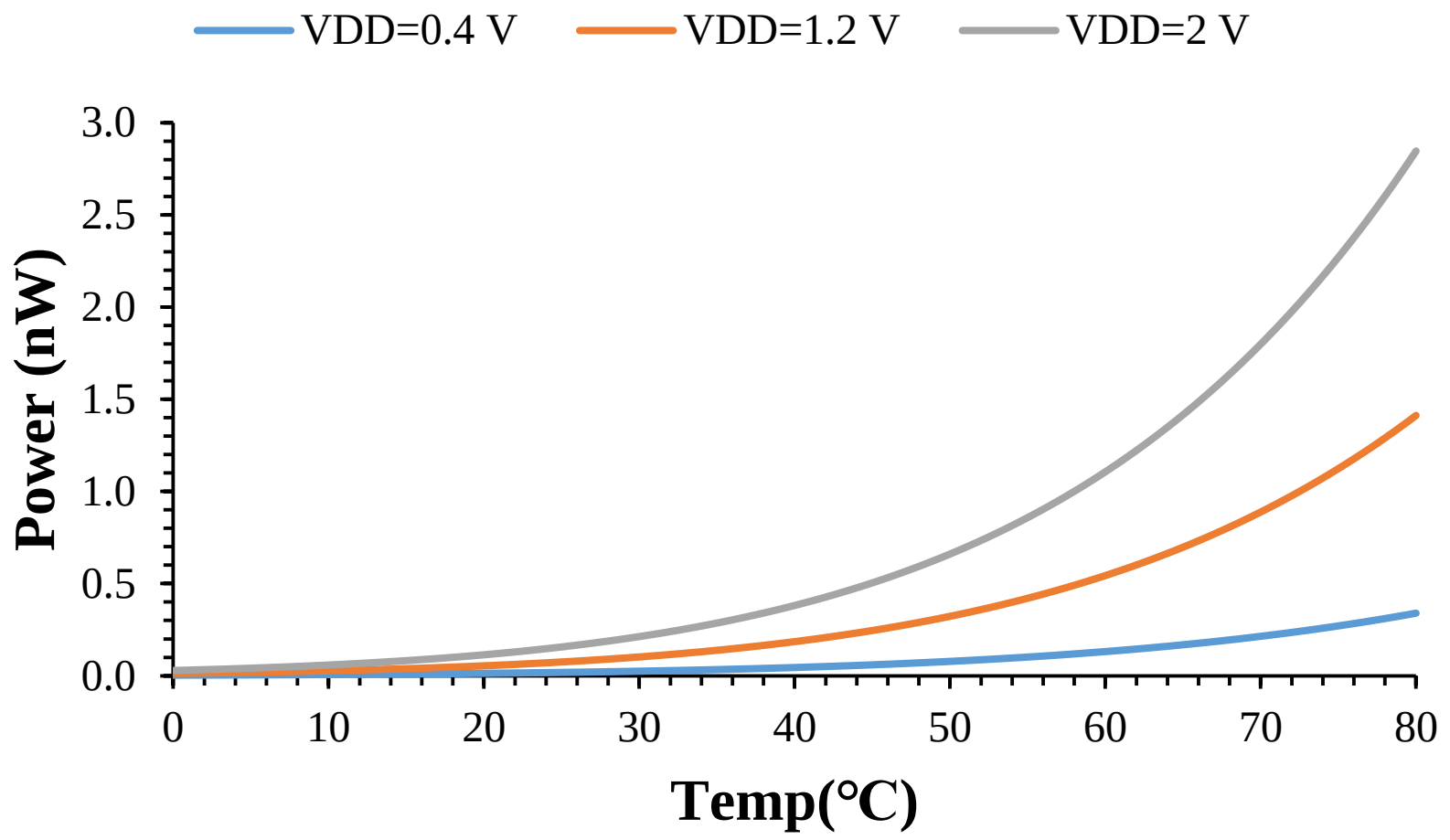

Figure 21- Changes in power consumption against temperature in 0.4 V, 1.2 V, and 2 V supply voltage

In order to study the process and mismatch changes, Monte Carlo analyses with 1000 runs on TC, LS, power consumption, and output reference voltage are executed and the results are compared with similar structures. Figure 22 shows the Monte Carlo analysis of the TC in the 0.4 V, 1.2 V, and 2 V supply potentials to examine the TC changes with the supply voltage. The average TCs are 40.41 ppm/°C, 40.44 ppm/°C, and 40.35 ppm/°C, respectively. Therefore, as expected, in the proposed design changes in TC are substantially low when the supply voltage is varied. Figure 23 shows the Monte Carlo analysis of LS, power consumption, and reference voltage at 25 °C which are 142.6 ppm/V, 19.2 pW, and 119.2 mV respectively. Figure 24 and Figure 25 present post-layout analyses to ensure design validity. The results of the schematic and post-layout analyses are compared in Table 2. The results of both analyses are relatively similar.

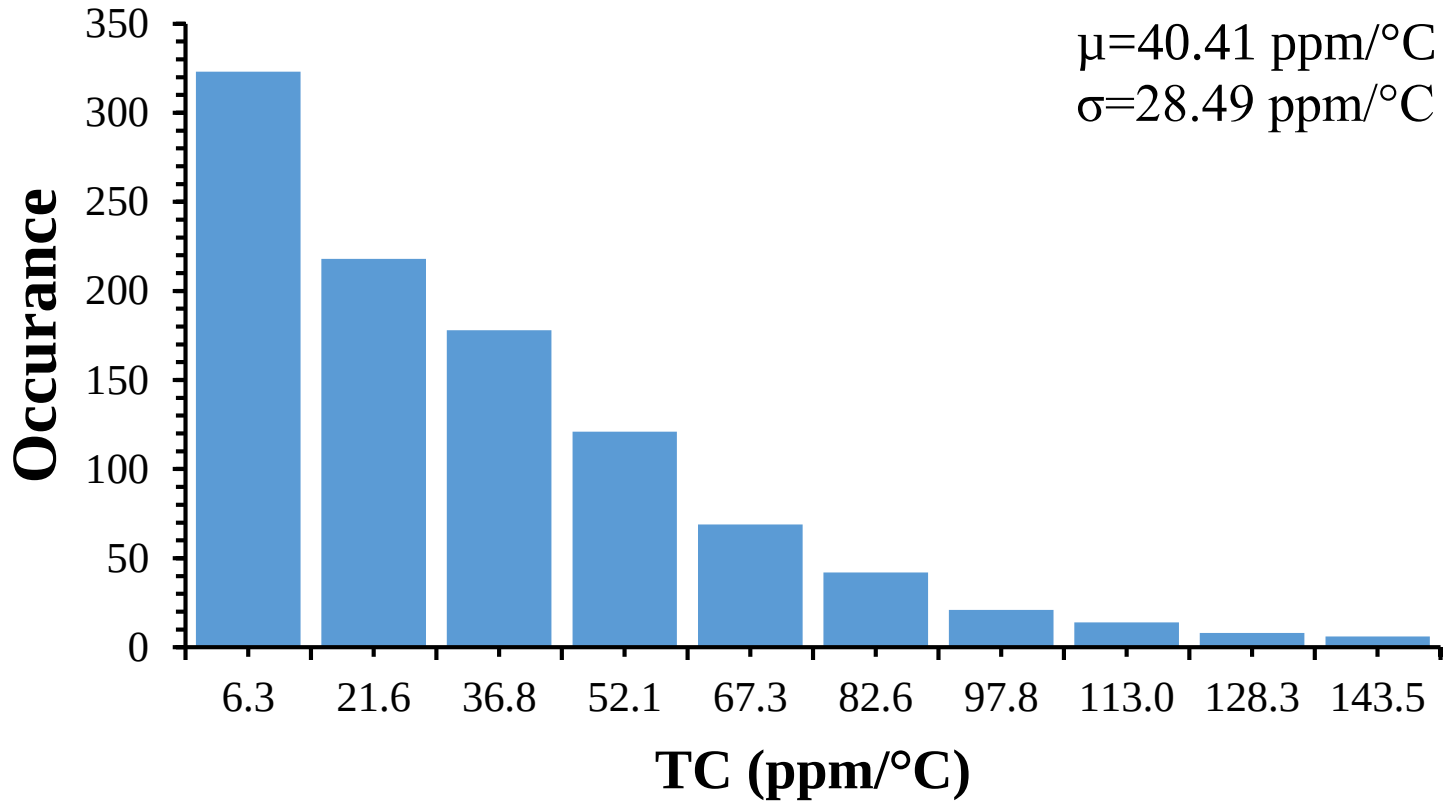

(a)

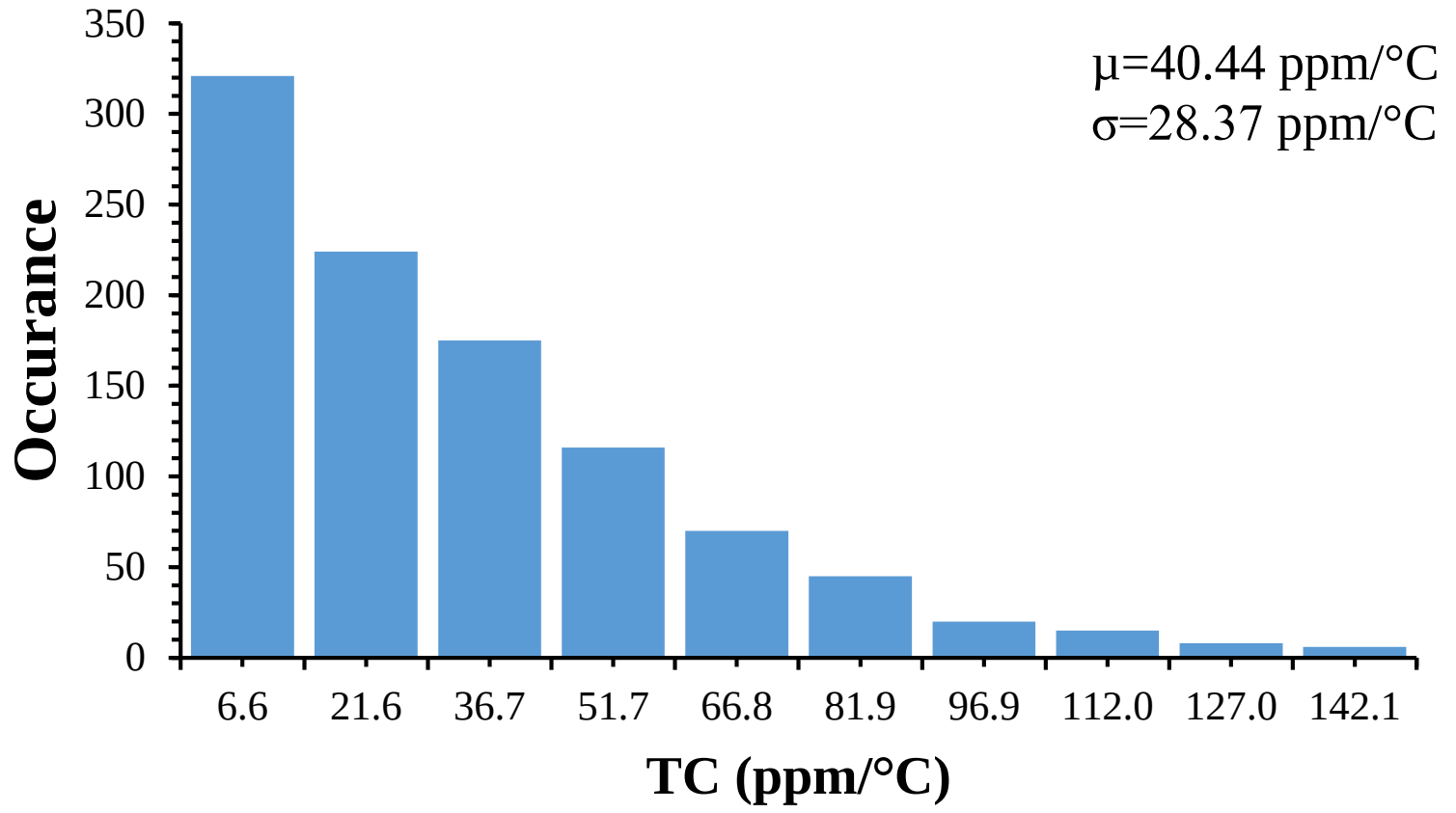

(b)

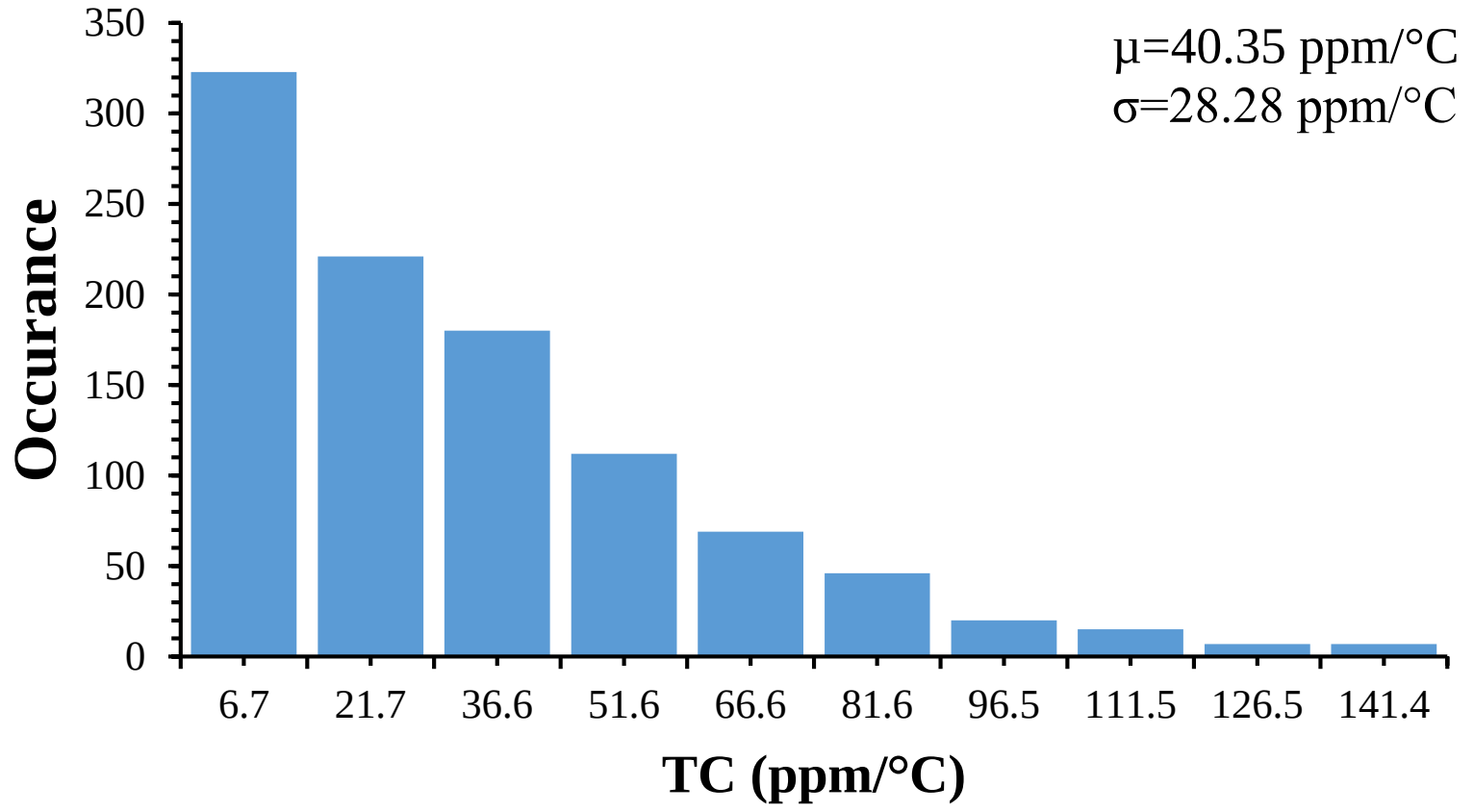

(c)

Figure 22- Monte Carlo analysis on the schematic circuit with 1000 runs on TC and with a supply voltage of (a) 0.4 V (b) 1.2 V (c) 2 V

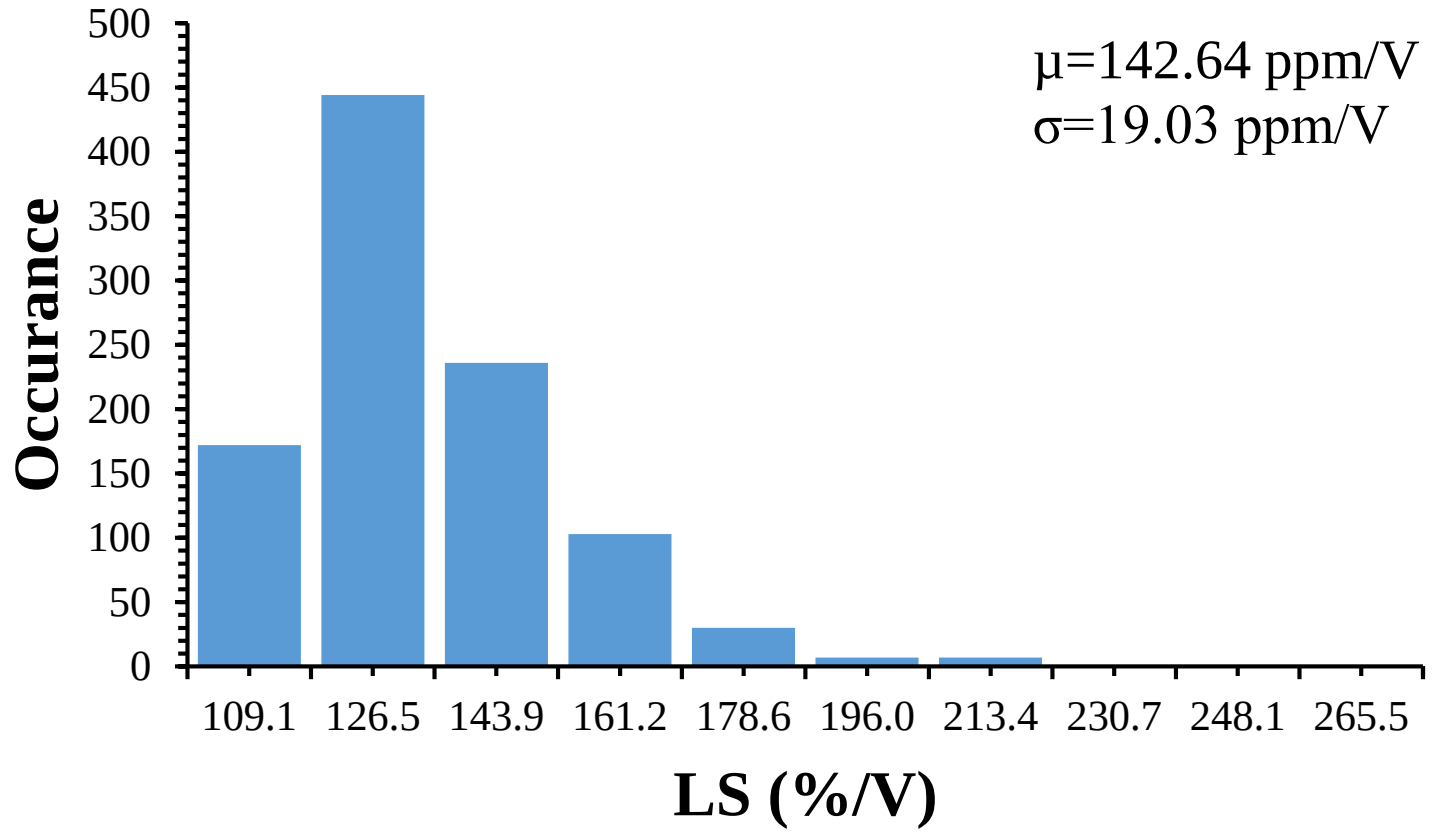

(a)

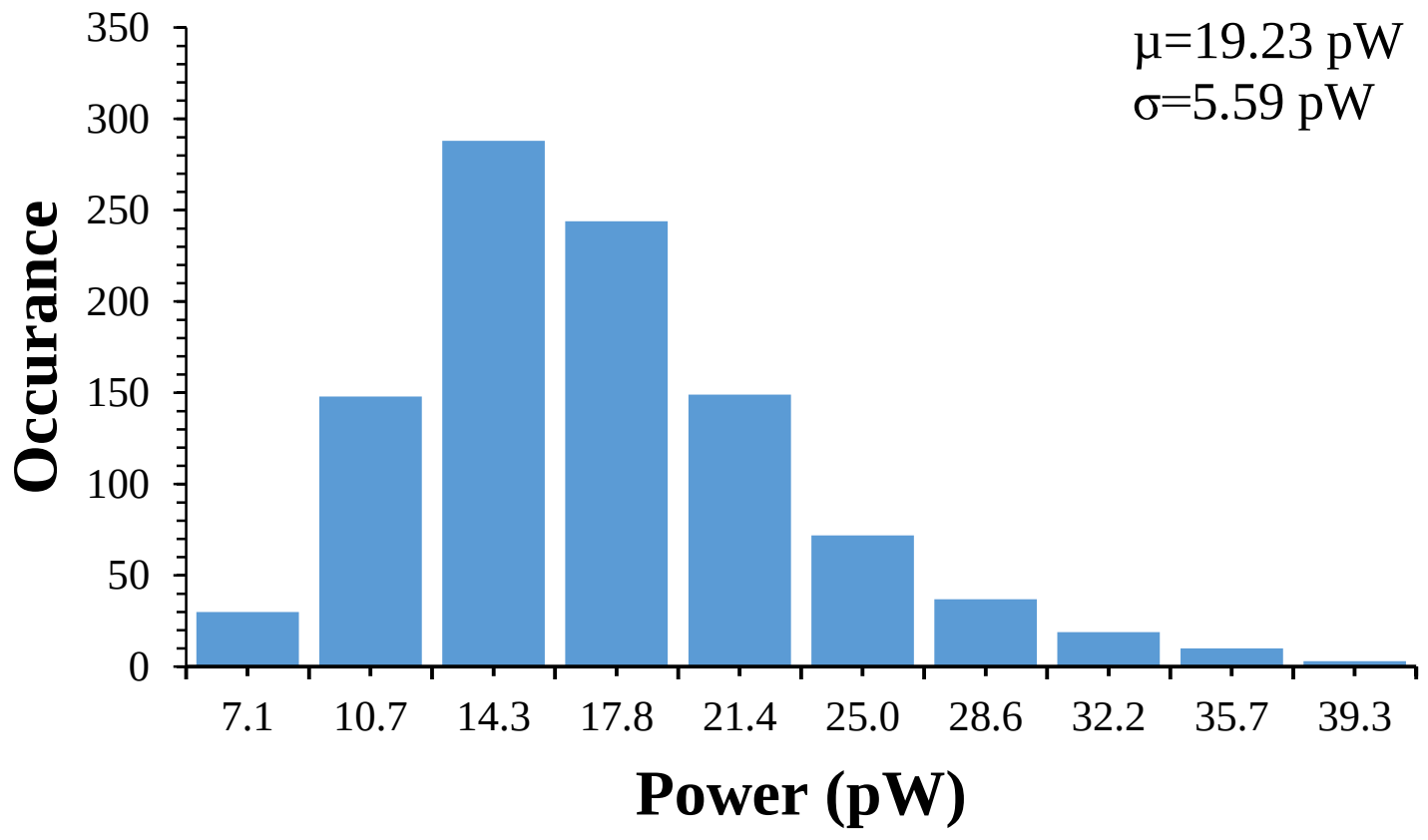

(b)

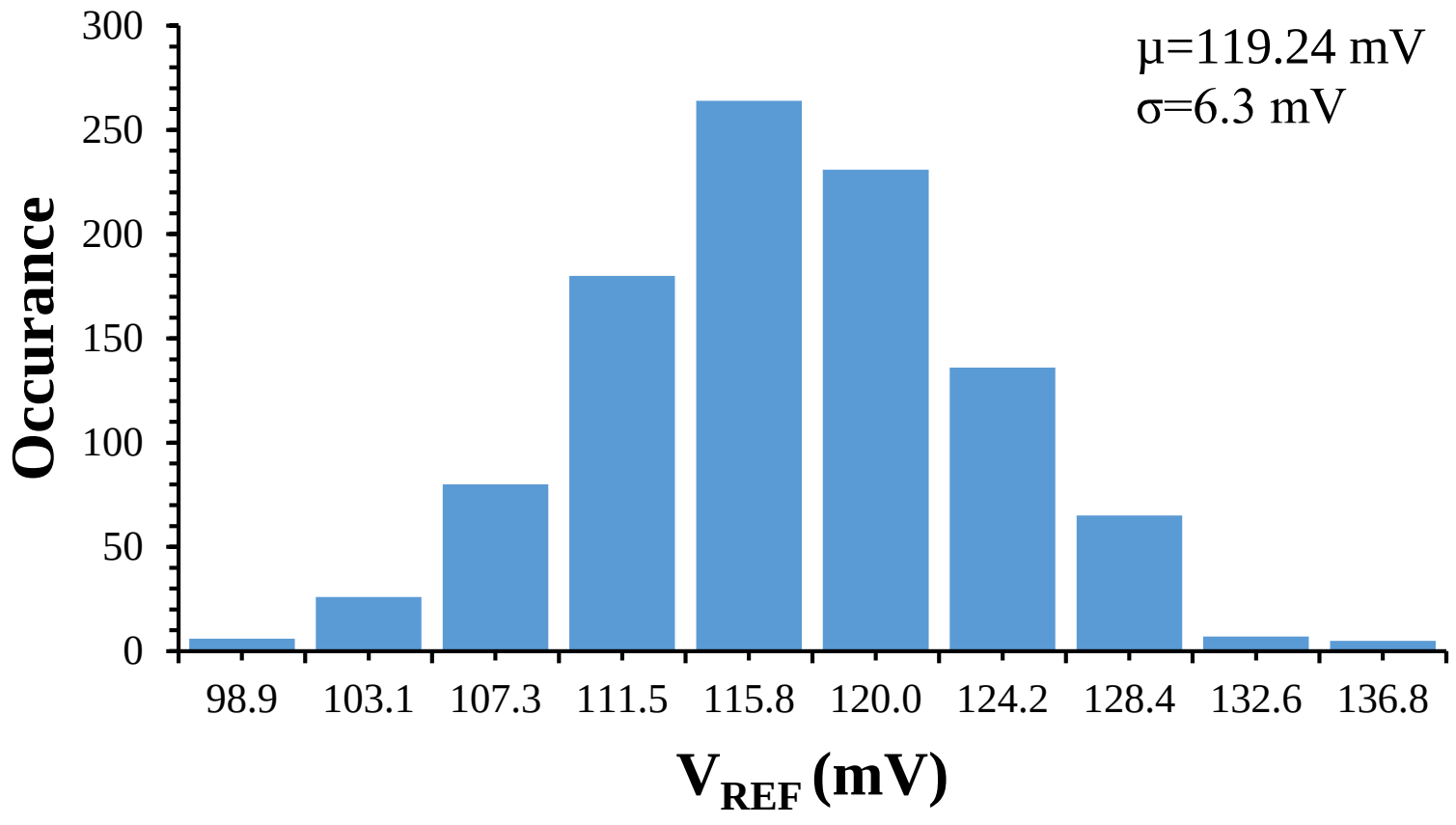

(c)

Figure 23- Monte Carlo analysis on the schematic circuit with 1000 runs at 25ºC. (a) LS (b) power consumption (c) output reference voltage

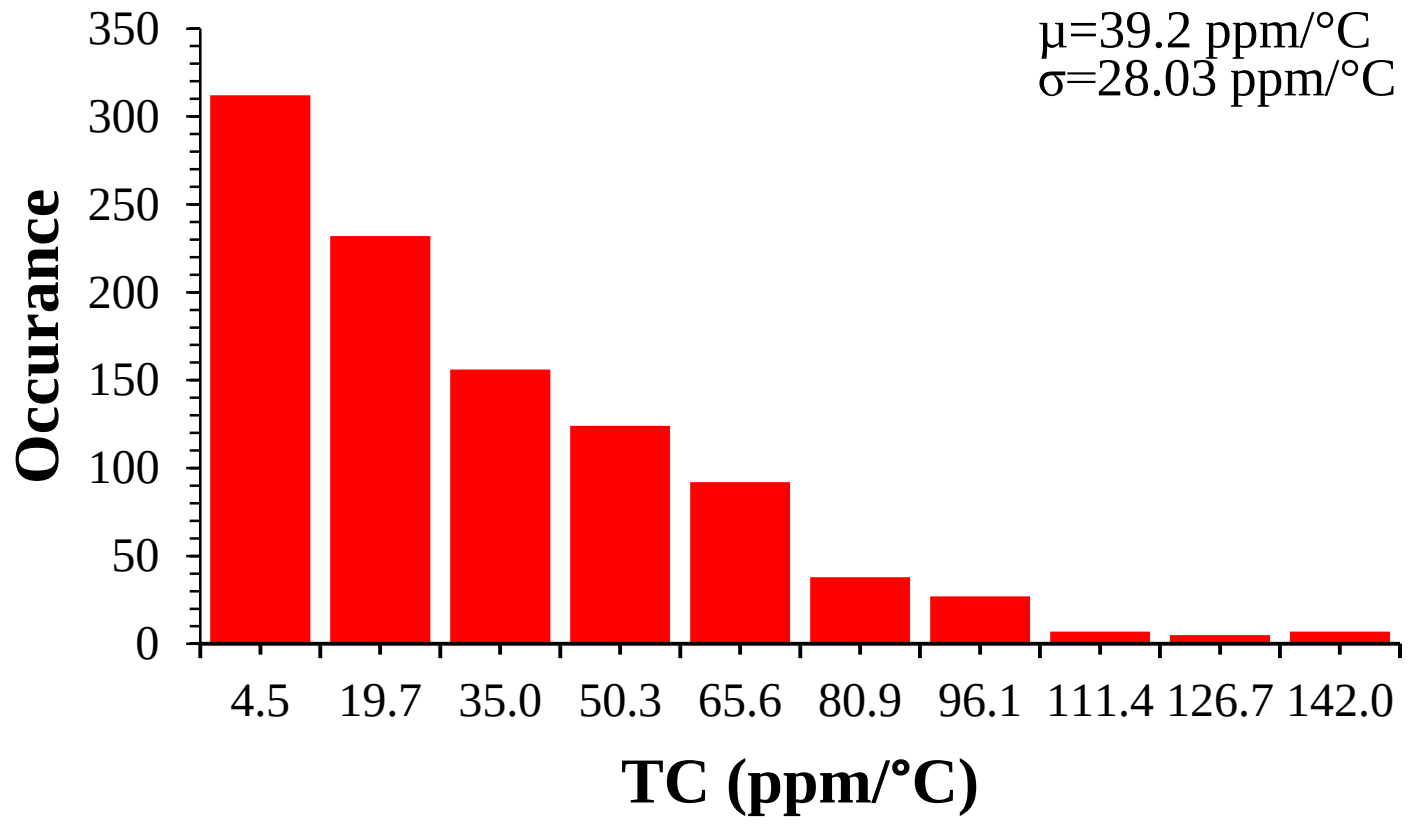

(a)

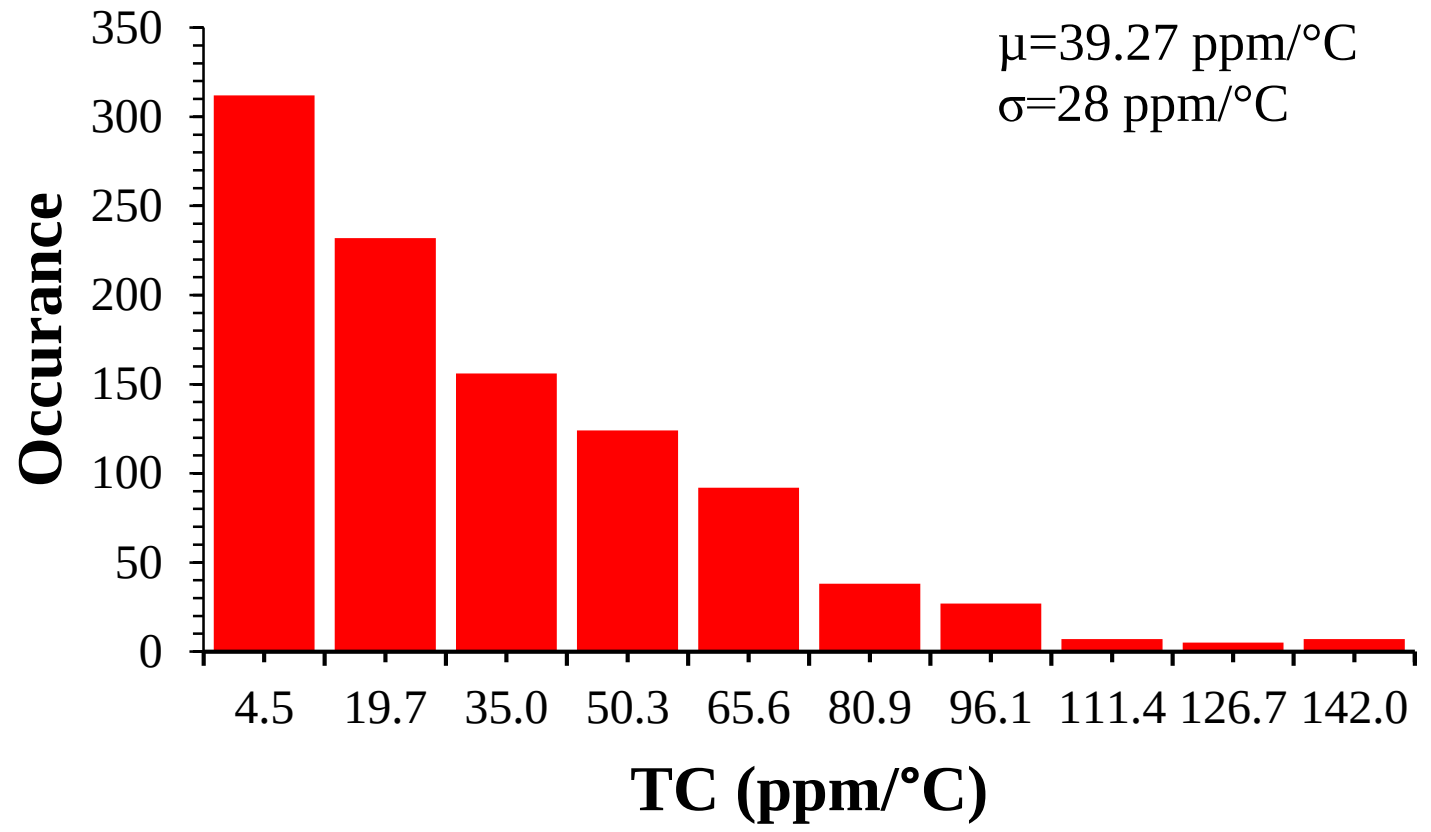

(b)

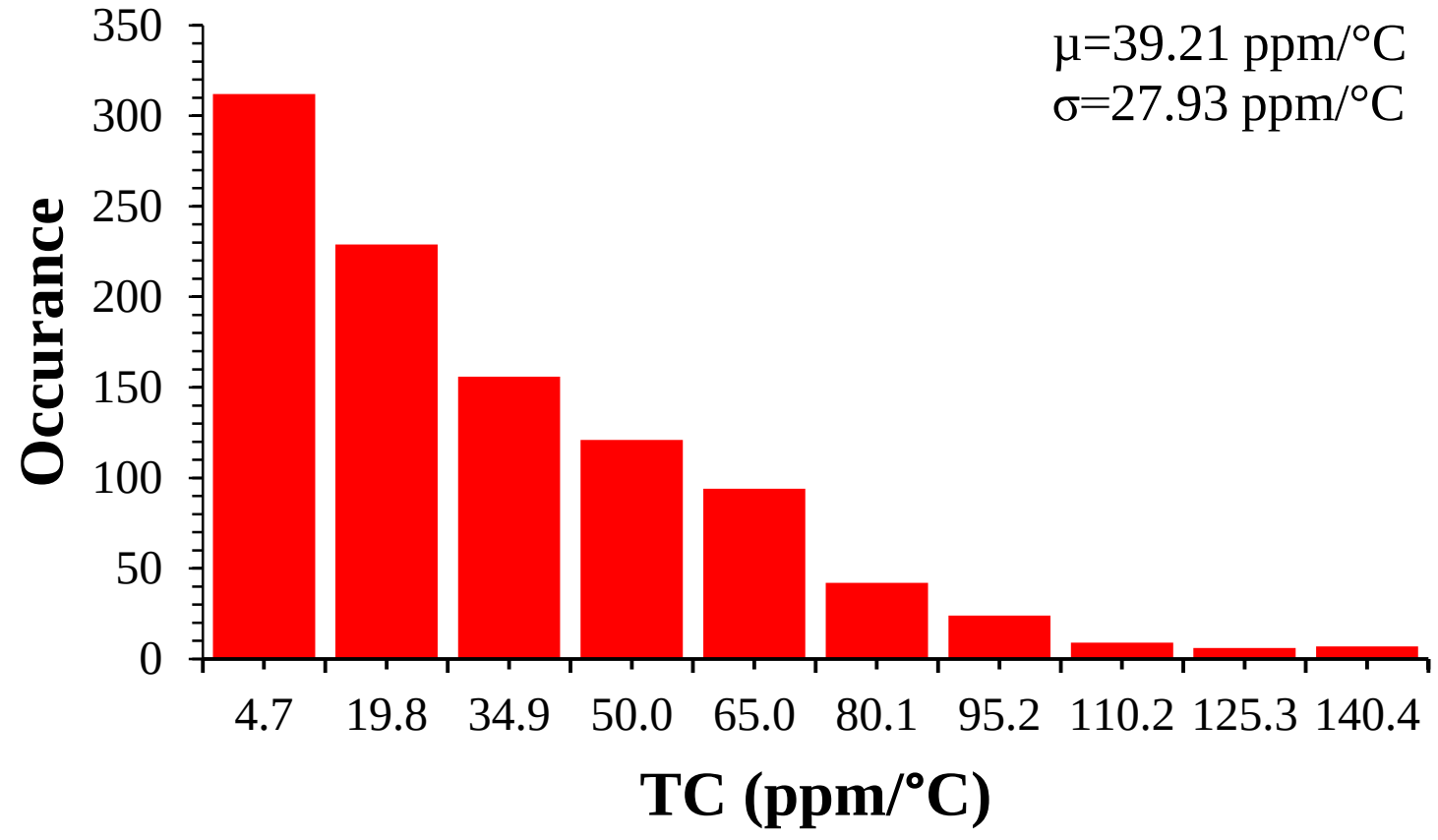

(c)

Figure 24- Monte Carlo analysis on the post-layout circuit with 1000 runs on TC and with a supply voltage of (a) 0.4 V (b) 1.2 V (c) 2 V

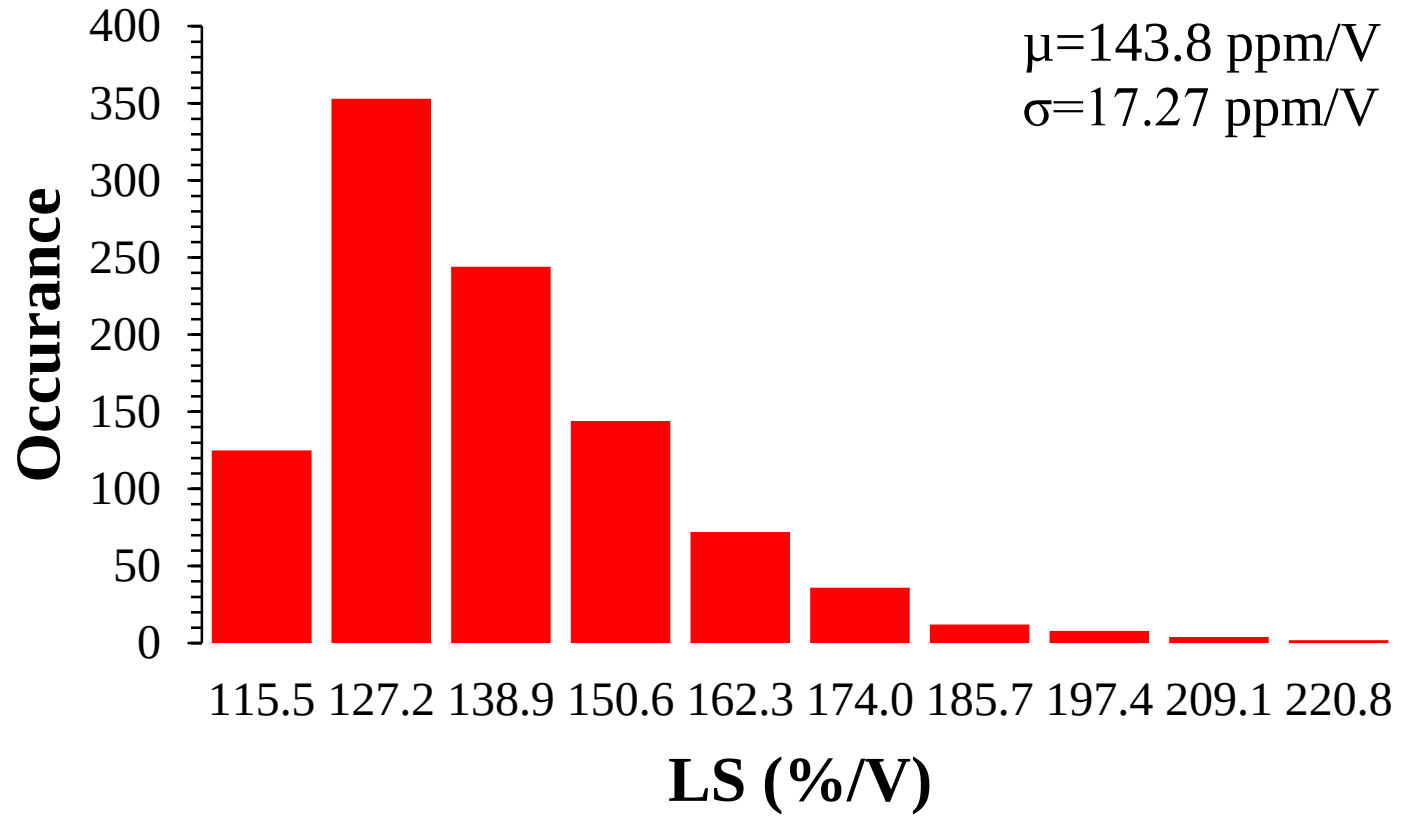

(a)

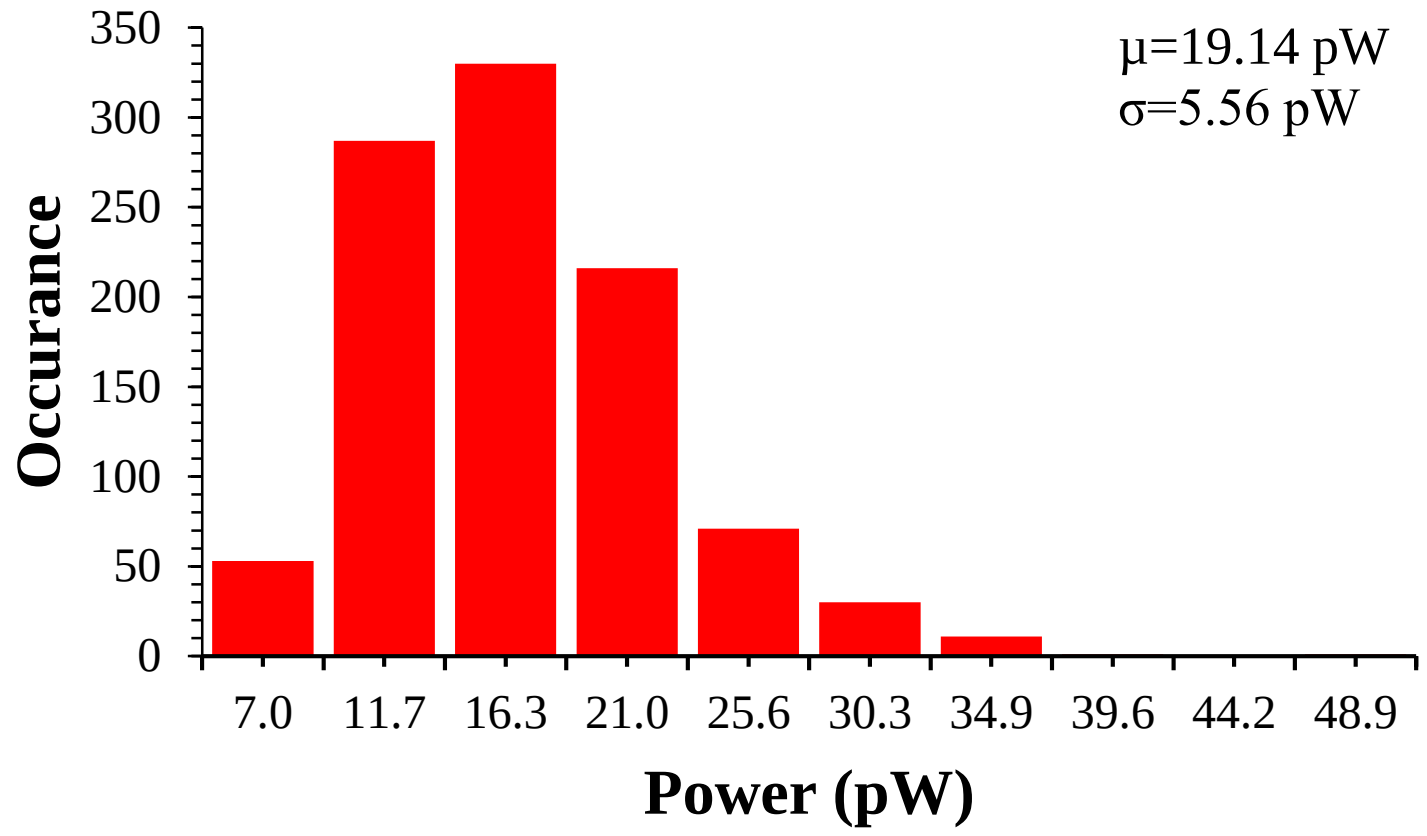

(b)

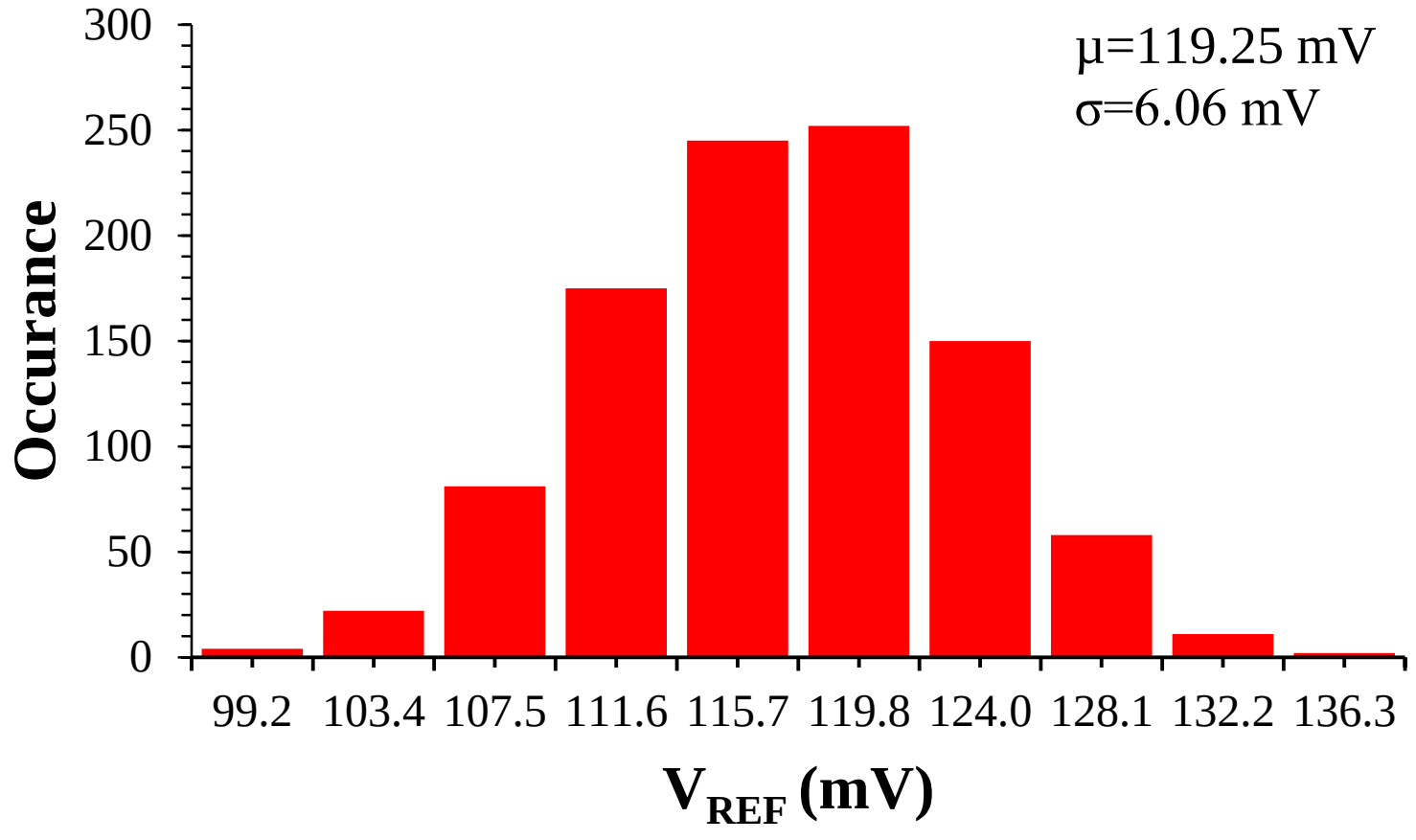

(c)

Figure 25- Monte Carlo analysis on the post-layout design with 1000 runs at 25ºC. (a) LS (b) power consumption (c) output reference voltage

Table 2- Comparison of Mont Carlo analysis between schematic and post-layout simulations with 1000 runs

| *Analysis* | *LS (ppm/V)* | | *TC (ppm/ºC)* | | *Power (pW)* | | *$V_{REF}$ (mV)* | |
|---|---|---|---|---|---|---|---|---|
| | μ | σ | μ | σ | μ | σ | μ | σ |
| *Schematic* | 142.6 | 19 | 40.4 | 28.5 | 19.2 | 5.6 | 119.2 | 6.3 |
| *Post-layout* | 143.8 | 17.3 | 39.2 | 28 | 19.1 | 5.6 | 119.2 | 6.1 |

Figure 26 shows the amount of output noise from 0.001 Hz to 100 Hz with no load capacitance. Increasing the load capacitance is a solution to reduce the amount of output noise.

Table 3 compares the proposed design with recent low power voltage references which have comparable values of LS and TC. The proposed voltage reference has the best LS and PSRR compared to other designs. The power consumption of the proposed design is among the lowest.

Although [11] has a lower power consumption without the use of trim, the LS and TC are worse than the proposed design and requires a higher operational supply voltage. Despite not using temperature trim, the TC of the proposed voltage reference has an acceptable value compared to similar references.

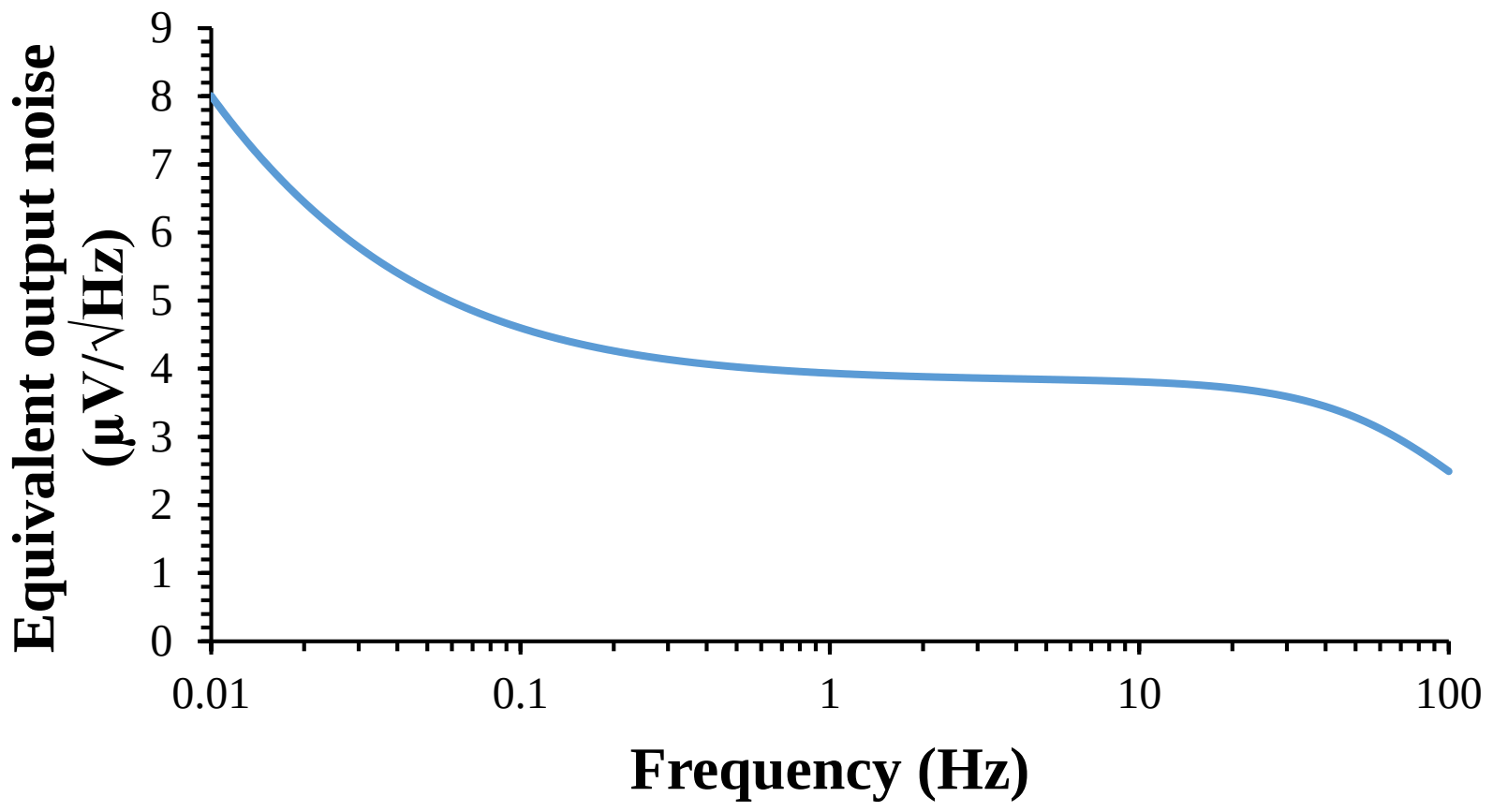

Figure 26- Output noise spectrum from 0.01 Hz to 100 Hz

Table 3- Comparison of the proposed design with recent similar voltage references

| **Design** | **This work*** | [12] | [9] | [22] | [16] | [15] | [30] | [11] | [14]* |
|---|---|---|---|---|---|---|---|---|---|
| **Tech (µm)** | 0.18 | 0.18 | 0.18 | 0.18 | 0.18 | 0.13 | 0.18 | 0.13 | 0.18 |
| **Min Supply (V)** | 0.4 | 0.45 | 1 | 0.34 | 0.45 | 0.4 | 1.4 | 0.5 | 0.2 |
| **LS (ppm/V)** | 143.8 | 4600 | 200 | 190 | 1500 | 800 | 3100 | 360 | 2799 |
| **Power (pW)** | 19.1 | 40 | 192 | 48$^T$ | 54.8/ 147$^T$ | 50.4 | 33.6 | 2.2 29.5$^T$ | 41.4 |
| **Temp range (°C)** | 0-80 | 0-120 | -20-100 | 0-100 | 0-120 | -25-125 | 0-100 | -20-80 | 10-70 |
| **TC (ppm/°C)** | 39.2 20.8$^T$ | 105.4 | 33 | 14.8$^T$ | 104/ 72.4$^T$ | 159 | 23 | 62 29$^T$ | 271.3 |
| **PSRR (dB) @10 Hz @10 KHz** | -90.9 -78 | N/A -29.2 | -62.5 N/A | -62.7 -50.2 | -44.3 -50.3 | N/A N/A | 42.24 -42.58 | -50.5 -58.5 | -57.1 -46.7 |
| **$V_{REF}$ (mV)** | 119.2 | 275.4 | 692.6 | 147.9$^T$ | 225.2/ 256.6$^T$ | 27.2 | 1250 | 176 | 21.16 |
| **Area (µm$^2$)** | 2183 | 18000 | 4500 | 33200 | 2000 | 1200 | 2500 | 9300 | 2149 |

*post-layout simulation
$^T$after trim

# 8- Conclusion

The voltage reference presented in this paper consists of two correction stages of LS and TC. The use of a LS corrector on the first stage together with a new DIBL compensation circuit improves the LS by nearly 97.6 percent. Multiplying the LS of the two stages, the average LS of the achieved reference voltage generator reaches 143.8 ppm/V. In addition, the LS correction stage dramatically improves the PSRR and reduces it by 38.3 dB and 25.4 dB at 10 Hz and 10 KHz respectively. To implement the proposed voltage reference, a 0.18 µm standard CMOS technology has been used, while process variation effects have been considered by applying Monte Carlo analyses on the schematic and post-layout extraction of the proposed circuit with 1000 runs. The minimum supply voltage required for this structure is 0.4 V which results in power consumption of 19.1 pW, one of the lowest values in recent designs. By changing the temperature from 0 ºC to 80 ºC, a TC value of 39.2 ppm/ºC is obtained. In general, the presented voltage reference generator benefits from good independency from the supply voltage and its fluctuations, which with its low power consumption and suitable TC makes it a suitable choice for IoT applications.